\documentclass[pre,twocolumn,showpacs,floatfix]{revtex4}
\usepackage{graphicx}
\usepackage{amsmath}
\usepackage{amssymb}
\usepackage{subfigure}
\newcommand{\ww}[1]{\underline{\underline{{\bf #1}}}}

\newcommand{\be}{\begin{equation}}
\newcommand{\ee}{\end{equation}}
\newcommand{\bea}{\begin{eqnarray}}
\newcommand{\eea}{\end{eqnarray}}
\newcommand{\ba}{\begin{aligned}}
\newcommand{\ea}{\end{aligned}}
\newcommand{\bma}{\begin{bmatrix}}
\newcommand{\ema}{\end{bmatrix}}
\newcommand{\Pp}{{\mathcal P}}
\newcommand{\kk}{{\ww{\mathcal K}}}

\newcommand{\EE}{{\mathcal E}}
\newcommand{\fext}{{\bf F}^{\text{ext}}}
\newcommand{\sti }{\ww{K}}
\newcommand{\stia }{\sti^{(1)}}
\newcommand{\stib }{\sti^{(2)}}
\newcommand{\trs}[1]{\hspace{.05em}\,^{\bf T}\hspace{-.1em}\ww{#1}}
\newcommand{\rig }{\ww{G}}
\newcommand{\rigt }{\trs{G}}
\newcommand{\norm}[1]{\vert \vert {\bf #1}\vert \vert}
\newcommand{\normm}[1]{\vert \vert #1\vert \vert}
\newcommand{\bi}{\begin{itemize}}
\newcommand{\ei}{\end{itemize}}
\newcommand{\im}{\item}
\newcommand{\nij}{{\bf n}_{ij}}

\newcommand{\Tij}{{\bf T}_{ij}}
\newcommand{\Fij}{{\bf F}_{ij}}
\newcommand{\rij}{{\bf r}_{ij}}
\newcommand{\hhh}{\textsc{h}}

\newcommand{\pk}{\ww{\pi}}
\newcommand{\cc}{\ww{\ww{C}}}
\newcommand{\wF}{\ww{F}}

\newcommand{\tr}{\text{tr}\hspace{.05em}}
\newcommand{\disty}{\displaystyle}
\newcommand{\ave}[1]{\langle #1 \rangle}
\begin{document}

\title{Internal states of model isotropic granular packings.\\
III. Elastic properties.}

\author{Ivana Agnolin\footnote{Present address: Geoforschungzentrum, Haus D, Telegrafenberg, D-14473 Potsdam Germany}
}
\author{Jean-No\"el Roux}
\email{jean-noel.roux@lcpc.fr}
\affiliation{Laboratoire des Mat\'eriaux et des Structures du G\'enie Civil\footnote{
LMSGC is a joint laboratory depending on Laboratoire Central des Ponts et Chauss\'ees, \'Ecole Nationale
des Ponts et Chauss\'ees and Centre National de la Recherche Scientifique},
Institut Navier, 2 all\'ee Kepler, Cit\'e Descartes, 77420 Champs-sur-Marne, France}

\date{\today}

\begin{abstract}
In this third and final paper of a series, elastic properties of numerically simulated isotropic packings of spherical beads
assembled by different procedures, as described in the first companion paper, and then subjected to a varying confining pressure, 
as reported in the second companion paper, are investigated.
In addition to the pressure, which determines the stiffness of contacts because of Hertz's law, elastic moduli are
chiefly sensitive to the coordination number $z$, which should not be regarded as a function of the packing density. 
Comparisons of numerical and experimental results for glass beads in the
$10\mbox{kPa}-10\mbox{MPa}$ pressure range reveal similar differences between dry samples prepared in a dense state by vibrations
and lubricated packings, so that the greater stiffness of the latter, in spite of their lower density, can be 
attributed to a larger coordination number. Effective medium type
approaches, or Voigt and Reuss bounds, provide good estimates of bulk modulus $B$, which can be accurately bracketed, but badly fail
 for shear modulus $G$, especially in low $z$ configurations under low pressure. This is due to the different response of tenuous, 
fragile networks to changes in load \emph{direction}, as compared
to load \emph{intensity}. In poorly coordinated packings, the shear modulus, normalized by the average contact
stiffness, tends to vary proportionally to the degree of force indeterminacy per unit volume, even though this quantity does not
vanish in the rigid limit.
The elastic range extends to small strain intervals and compares well with experimental observations on sands. The origins of nonelastic 
response are discussed.
We conclude that elastic moduli provide access to mechanically 
important information about coordination numbers,
which escape direct measurement techniques, and indicate further perspectives.
\end{abstract}

% 45.70.-n      Granular systems
% 46.65.+g      Random phenomena and media

\maketitle

\section{Introduction}
The mechanical properties of granular materials and their relations to the packing microstructure are currently being investigated by many research groups.
As a simple model, long studied for its geometric aspects~\cite{CC87,BD91}, the packing of 
 equal-sized spherical balls is also mechanically characterized in the laboratory~\cite{DM57,KJ02,gilles03,Makse04,DO77}, and by numerical 
means, relying on discrete, granular level modeling~\cite{TH00,SUFL04,MGJS99,SEGHL02,SGL02,Makse04}.

The present paper is the last one in a series of three, about geometric and mechanical properties of bead packings obtained by numerical simulations. It 
focusses on the elastic properties of isotropically compressed samples. The study is based on the configurations for which the packing processes and resulting
microstructure were studied in paper I~\cite{iviso1}, while paper II~\cite{iviso2} reported on the effects of isotropic compressions and pressure cycles.
The results presented here, although perhaps better appreciated on knowing the contents of papers I and II, can be understood without reading these previous
contributions.
  
Elastic properties of granular assemblies are probed when small stress increments
are superimposed on a prestressed equilibrium configuration, either on 
controlling very small strains in a static experiment~\cite{Tat104,HI96,KJ02,GBDC03}
or in dynamical ones, relying
on resonance modes~\cite{CIJ88a,CIJ88b,THHR90}, or sound propagation~\cite{THHR90,Tat104,jia99,JM01,gilles03,KJ02,GBDC03,SDH04,ARMJM05}.
Elastic behavior of granular materials is only applicable for very small strain increments, typically
of order $10^{-5}$ or even $10^{-6}$ in usual conditions, \emph{i.e.,} with sands under 
confining stresses between 10~kPa and a few MPa~\cite{Tat104,HI96,KJ02,GBDC03}. It has been checked in such cases that
static measurements of elastic moduli, with devices accurate enough to control such small 
strains are consistent with ``dynamical'' ones, \emph{i.e.} deduced from 
experiments on wave propagation or resonance frequencies. Experimental soil mechanics have achieved a high 
level of sophistication, with significant progress over the last
twenty years~\cite{TAT01,Lyon03b}, and accurate measurements of the mechanical response of granular 
materials in the very small strain r\'egime are one example thereof. Coincidence of
elastic moduli values obtained by different means is reported, \emph{e.g.}, in~\cite{THHR90,Tat104,GBDC03}. 

The elastic moduli should not be confused with the slopes of stress-strain curves
on the scale of the strain level (usually in the 1\% range) corresponding to the full mobilization of internal friction. Such slopes are considerably smaller
than true elastic moduli (by more than an order of magnitude), and do not correspond to an elastic, reversible response. In this respect, the frequent use, 
for engineering applications, of a simplified elastoplastic behavior, in which the material is linear elastic until 
the Mohr-Coulomb criterion for plasticity is reached,  
as presented in Ref.~\cite{VER98}, should not be misinterpreted. Such crude
models, in which strains are elastic and reversible up to the complete mobilization of internal friction, 
are resorted to in engineering practice when detailed information on the constitutive law are not available, but the 
``elastic moduli'' introduced in those simplified constitutive relations are merely convenient parameters enabling one to perform approximate calculations. 

In micromechanical~\cite{WAK87,JimLuigiflu05} and numerical~\cite{MGJS99,Makse04,SRSvHvS05} studies, 
elastic properties are associated with the deformations of a fixed contact network, and should therefore correspond to the ``true elastic''
behavior observed in the laboratory for very small strain intervals. Indeed, except in very special
situations in which the effects of friction are suppressed and geometric restructuration is reversible~\cite{JNR97b,JNR2000}, 
the irreversible changes associated with network alterations or rearrangements preclude all kind of elastic modelling. 
Elastic properties are attached to one specific contact set, and hence of limited relevance
to the rheology of granular materials.
Nervertheless, elastic properties are interesting because they might provide access, 
in a non-destructive way, to geometric data on the contact network, such as coordination numbers. Such variables
are still virtually inaccessible to direct measurements, 
even with sophisticated visualization techniques, as emphasized in paper I~\cite{iviso1}, but they are very likely, in turn, to influence the
constitutive laws for larger strains.

This paper is organized in the following way. We first recall the properties of the model material we are studying (Section~\ref{sec:model}), along with basic definitions
and properties pertaining to the elasticity of granular packings. Then, useful results on the pressure-dependent internal states of the various types of configurations
introduced and studied in papers I and II~\cite{iviso1,iviso2} are summarized in Section~\ref{sec:assemb}.
Next, the values of elastic constants in the different configuration series, 
as a function of (isotropic) confining pressure, are presented in Section~\ref{sec:elastp}, where their relations to internal structure  are also discussed. 
Section~\ref{sec:elast0} is devoted to the particular behavior of elastic moduli in the tenuous contact networks of poorly coordinated configurations.
Numerical results are confronted to experimental ones in Section~\ref{sec:elastexp}. Some results about the extension of the elastic range
are given in Section~\ref{sec:elastreg}. Section~\ref{sec:conc} discusses the results and indicates some further perspectives.

\section{Numerical model and basic definitions\label{sec:model}}
Packings of $n$ spherical beads are simulated with molecular dynamics, in which equations of motion resulting from Newton's laws are solved for
the particle positions and rotations. 
Thanks to a suitably adapted form of the Parrinello-Rahman deformable cell molecular dynamics technique~\cite{PARA81,PR82}, 
as described in paper I~\cite{iviso1}, we request all three diagonal components of
the Cauchy stress tensor, denoted as $\sigma_{\alpha\alpha}$ to be equal, in equilibrium, 
to prescribed values $\Sigma_{\alpha}$, all chosen to coincide 
with a pressure $P$ in this study of isotropic states. Differences between $\sigma_{\alpha\alpha}$ and $\Sigma_{\alpha}$
entail some evolution in the cell size parameters.
$\sigma_{\alpha\alpha}$ is given in equilibrium by the classical formula:
\be
\sigma_{\alpha\alpha}=
 \frac{1}{\Omega}\sum_{i<j} F_{ij}^{(\alpha)} r_{ij} ^{(\alpha)},
\label{eqn:stress}
\ee
where the sum runs over all pairs in contact, $F_{ij}^{(\alpha)}$ is the $\alpha$ coordinate 
of the force $\Fij$ exerted by grain $i$ onto its neighbor $j$ at their contact, while
$r_{ij}^{(\alpha)}$ is the $\alpha$ coordinate of vector $\rij$, pointing from the center of $i$ to the center of $j$.
From Eqn.~\eqref{eqn:stress} one can easily deduce a simple and useful relation between pressure $P = (\sigma_{11}+ \sigma_{22}+ \sigma_{33})/3$ and the average normal
contact force $\ave{N}$ between monosized spheres of diameter $a$, involving solid fraction $\Phi$ and coordination number $z$~:
\be
P= \frac{z\Phi \ave{N}}{\pi a^2},
\label{eqn:relpn}
\ee
The corresponding dynamical equations used to impose stresses are described in paper I~\cite{iviso1}, and we briefly recall here the essential ingredients of the model
for a study of the elastic response of packings that have been first assembled and compressed, as reported in papers I and II. \emph{Dynamical} aspects of the model,
in particular (inertia, viscous dissipation) play no role in the determination of elastic moduli, for which our calculations are based on the building of
the \emph{stiffness matrix} of the contact network.  
\subsection{Local stiffnesses \label{sec:forces}}
We consider spherical beads of diameter $a$, with the elastic properties of glass: Young modulus $E=70$~GPa, Poisson ratio $\nu=0.3$. They interact  
in their contacts by the Hertz law, which states that the elastic normal force $N$ is proportional to $h^{3/2}$, $h$ being the normal deflection of the contact,
so that the incremental normal stiffness $\frac{dN}{dh}$, with the notation $\tilde E = \frac{E}{1-\nu^2}$, is given by:
\be
K_N = \frac{dN}{dh} = \frac{\tilde E\sqrt{a}}{2}h^{1/2} = \frac{3^{1/3}}{2} \tilde E^{2/3} a^{1/3} N^{1/3}
\label{eqn:kn}
\ee
The tangential elastic force is to be incrementally evaluated with a simplified Mindlin-Deresiewicz form~\cite{JO85} on assuming 
the tangential stiffness $K_T$ to stay proportional to $K_N$, and hence a function of $N$:
\be
K_T = \alpha _T  K_N \ \ \mbox{with}\ \ \alpha_T=\frac{2-2\nu}{2-\nu}.
\label{eqn:tang}
\ee
The tangential force ${\bf T}$ is constrainted by the Coulomb condition $\norm{T}\le \mu N$, with friction coefficient $\mu$ set to $0.3$, and additional 
conditions are introduced to ensure thermodynamic consistency and objectivity (see paper I~\cite[Sec. II, appendices A and B]{iviso1}).

All our results will be stated in a form independent of
bead diameter $a$.
All dimensionless results -- such as \emph{e.g.} ratios of elastic moduli of the granular material to $\tilde E$ -- depend, in addition to $\nu$, on
a reduced stiffness parameter we define as 
$$\kappa = \left(\frac{\tilde E}{P}\right)^{2/3}.$$
Typical values of contact deflections $h$ scale as $\kappa^{-1}a$.

Let  $N_c$ denote the number of force-carrying contacts. In every contacting pair $i$-$j$,
we arbitrarily choose a ``first'' grain $i$ 
and a ``second'' one $j$, and define the relative dispacement vector $\delta {\bf u}_{ij}$ as the difference between the displacement of the contact point
as belonging to solid $i$ and its displacement as a point belonging to solid $j$, both regarded as rigid.

In each contact the force ${\bf F}_{ij}$ that is transmitted from $i$ to $j$ is split into its normal
and tangential components as ${\bf F}_{ij} = N_{ij}{\bf n}_{ij} + {\bf T}_{ij}$. 
The static contact law 
relates the $3N_c$-dimensional
contact force increment vector $\Delta {\bf f}$, formed with the values
$\Delta N_{ij}$, $\Delta \Tij $ of the normal and tangential parts of all contact force increments, to 
$\delta {\bf u}$:
\be
\Delta {\bf f} = \kk \cdot \delta {\bf u}.
\label{eqn:cmat}
\ee
This defines the $(3N_c\times 3N_c)$ matrix of contact stiffnesses $\kk $. $\kk $ is block diagonal (it does not
couple different contacts), and we shall refer to it as the \emph{local stiffness matrix}. The $3\times 3$  block of $\kk$
corresponding to contact $i,j$, $\kk _{ij}$ is diagonal itself provided friction is
not fully mobilized, and contains stiffnesses $K_N(h_{ij})$ and (twice in 3 dimensions) $K_T(h_{ij})$ as
given by~\eqref{eqn:kn} and~\eqref{eqn:tang}:
\be
\kk _{ij}^E =\bma K_N(h_{ij}) & 0 & 0 \\ 0 &K_T(h_{ij})& 0 \\ 0&0&K_T(h_{ij}).
\ema
\label{eqn:kiju}
\ee
This simple form of $\kk _{ij}$ ignores some non-diagonal terms that appear when the normal force decreases, or for sliding contacts. The effects of those terms are 
discussed and tested in Appendix~\ref{sec:appendixfnft}, as well as the possible use of more elaborate contact laws~\cite{THRA88}, in which $K_T/K_N$ is not
kept constant.

\subsection{Global rigidity and stiffness matrices\label{sec:rist}}
When elastic properties are investigated, small displacements about an
equilibrium configuration are dealt with to first order (as an infinitesimal motion,
\emph{i.e.} just like velocities), and
related to small increments of applied forces, moments and stresses. We use periodic boundary conditions in our simulations, and 
the dimensions $L_\alpha$ ($\alpha=1$, $2$, $3$) of the parallelipipedic simulation cell are degrees of freedom of the system, while
all three diagonal components of the stress tensor are externally imposed~\cite{iviso1,iviso2}.
We use three strain parameters defined as the relative changes of those lengths, from their values in a reference state:
$$\epsilon_\alpha = -\Delta L_\alpha/L_\alpha$$
With our conventions, shrinking strains and compressive stresses are positive.

Let us now recall the definition of the \emph{rigidity matrix} (not to be confused with the stiffness matrix), as introduced in paper I.
The grain center displacements $({\bf u}_i)_{1\le i\le n}$ are
conveniently written as 
\be
{\bf u}_i = \tilde {\bf u}_i - \ww{\epsilon}\cdot {\bf r}_i,
\label{eqn:utilde}
\ee
with a set of displacements $\tilde {\bf u}_i$ satisfying periodic boundary conditions in the cell with the current 
dimensions, $\ww{\epsilon}$ denoting the diagonal strain matrix with coefficients $\epsilon _\alpha$. 
Gathering all coordinates of particle (periodic) displacements and rotation
increments, and strain parameters one defines a
\emph{displacement vector} in a space with dimension equal to the number of degrees of freedom $N_f=3n+3$ (recall $n$ is the number of beads),
\be
{\bf U} = \left( ( \tilde {\bf u}_i, \Delta {\bf \theta} _i)_{1\le i\le n},
(\epsilon_\alpha)_{1\le \alpha\le 3}\right).
\label{eqn:defu}
\ee

The normal unit vector $\nij$ points from $i$ to
$j$ (along the line joining centers for spheres).
The relative displacement $\delta {\bf u}_{ij}$, for spherical grains with radius $R$, reads
\be
\delta {\bf u}_{ij}= \tilde {\bf u}_i +\delta {\bf \theta}_i\times R \nij -\tilde {\bf u}_j
+\delta {\bf \theta}_j\times R \nij + \ww{\epsilon}\cdot {\bf r}_{ij}, \label{eqn:deprel1}
\ee
in which ${\bf r}_{ij}$ is the vector pointing from the center of the first sphere $i$ 
to the nearest image (by the periodic translation group of the boundary conditions) of the center of
the second one $j$. The normal part $\delta {\bf u}_{ij}^N$ of $\delta {\bf u}_{ij}$
is the increment of normal deflection $h_{ij}$ in the contact. 

The rigidity matrix $\ww{{\bf G}}$ is $3N_c\times N_f$-dimensional, it is defined by the linear correspondence expressed by relation~\eqref{eqn:deprel1},
which transforms ${\bf U}$ into the $3N_c$-dimensional
vector of relative displacements at contacts $\delta {\bf u}$:
\be
\delta {\bf u}=\rig \cdot {\bf U}
 \label{eqn:deprel}
\ee

External forces ${\bf F}_i$ and moments ${\bf \Gamma}_i$ 
applied to grain centers, and diagonal Cauchy stress components $\Sigma_\alpha$ can be gathered in one $N_f$-dimensional
\emph{load vector} $\fext$:
\be
\fext = \left(({\bf F}_i,{\bf \Gamma}_i)_{1\le i\le n},(\Omega \Sigma_\alpha)_{1\le\alpha\le 3} \right),
\label{eqn:deffext}
\ee
chosen such that the work in a small motion is equal to $\fext \cdot {\bf U}$.
The equilibrium equations -- the statements that contact forces ${\bf f}$ balance load $\fext$ --
is simply written with the tranposed rigidity matrix, as
\be
\fext = \rigt\cdot {\bf f}.
\label{eqn:trig}
\ee
This is of course easily checked on writing down all force and moment coordinates, as well as the equilibrium
form of stresses, Eqn.~\ref{eqn:stress} with $\sigma_{\alpha \alpha}=\Sigma_{\alpha} $ for all $\alpha$.
%\be
%\Omega \Sigma_\alpha = \sum _{i<j} F_{ij} ^\alpha  r_{ij}^\alpha.
%\label{eqn:stressequil}
%\ee

Given \eqref{eqn:deprel}, which defines the rigidity matrix, relation \eqref{eqn:trig} is a statement of the theorem of virtual work (see paper I~\cite{iviso1}).

Returning to the case of small displacements associated with a load \emph{increment} $\Delta \fext$, one may write,
to first order in ${\bf U}$,
\be
\Delta \fext = \sti \cdot {\bf U},
\label{eqn:defsti}
\ee
with a \emph{total stiffness matrix} $\sti$, comprising two parts, $\stia$ and $\stib$, which we respectively
refer to as the \emph{constitutive}
and \emph{geometric} stiffness matrices. $\stia$ results from
Eqns.~\ref{eqn:deprel}, \ref{eqn:cmat} and \ref{eqn:trig}
\be
\stia = \rigt\cdot \kk \cdot \rig
\label{eqn:sti1}
\ee
$\stib$ is due to the change of the geometry of the packing, and is written down in
Appendix~\ref{sec:appendixTrot}, where it is also shown to be negligible in general.

To the rigidity matrix are associated the concepts  of force and velocity (or
displacement) indeterminacy, of relative displacement compatibility and of
static admissibility of contact forces. Those definitions are given in paper I, where they are used to discuss 
the limit of isostatic packings~\cite{JNR2000}.

\subsection{Grain-level and macroscopic elasticity \label{sec:defelastic}}
We now discuss the conditions for which the response to load increments of a 
prestressed granular packing in mechanical equilibrium can be described as elastic, and explain how 
macroscopic elastic moduli are computed in our simulations.
\subsubsection{Some necessary approximations}
Elasticity implies the existence of an elastic potential (a function of ${\bf U}$) from which forces are derived.
If force increments are written as linearly depending on displacements, as in~\eqref{eqn:defsti}, the corresponding
stiffness matrix $\sti$ should be unique (the same for all ${\bf U}$ vectors) and symmetric. Strictly speaking, the
differences in the form of the local stiffness matrix $\kk$ on the direction of displacements
precludes the definition of an elastic response. 
This is discussed in Appendix~\ref{sec:appendixfnft}. We could 
check that this effect is quantitatively negligible, and that taking all $\kk _{ij}$
block values as in Eqn.~\eqref{eqn:kiju} is a very good approximation. Likewise, the implementation of more
complicated contact laws accounting for a gradual change of tangential stiffness $K_T$ as a function of
$\norm{T}$ (see Appendix~\ref{sec:appendixfnft}), entails negligible changes in the values of elastic moduli.
 
Moreover, the non-symmetric geometric contribution $\stib$  is also negligible -- provided simple
precautions are applied to prevent the free motion of divalent beads, 
as explained in Appendix~\ref{sec:appendixTrot}. These motions, as checked in paper I~\cite{iviso1}, do not jeopardize global stability, and they are
the only \emph{mechanisms } (\emph{i.e.,} non-zero elements of the kernel of rigidity matrix $\rig$) of the backbone. 
On using the symmetric diagonal
form~\eqref{eqn:kiju} for all contacts, $\kk$ will be symmetric and positive definite. In view of~\eqref{eqn:sti1}, 
the kernel of $\stia$ (the ``floppy modes'' of the constitutive stiffness matrix) 
coincides with the kernel of $\rig$ (the ``mechanisms''). A suitable elimination of the localized free motion of 2-coordinated grains
(see Appendix~\ref{sec:appendixTrot}) enables one to work with a positive definite global stiffness matrix, 
and therefore with a well-behaved elastic potential energy. The elastic regime, however,
is only defined on \emph{approximating} the contact laws as elastic.

Finally, as explained in Appendix~\ref{sec:appendixcc}, initial confining stresses entail
very small corrections to elastic moduli, of order $P$, which we also neglect.

\subsubsection{Computation of elastic moduli\label{sec:calcul}}
In order to evaluate macroscopic
elastic moduli or compliances, one can apply stress increments and measure the resulting strains. With our choice of boundary
conditions and degrees of freedom, we choose load increments $\Delta \fext$ with all coordinates set to zero except one of the three
last ones, say $\Omega \Delta \sigma_{\alpha\alpha}$, corresponding to a diagonal stress increment, according
to definition~\eqref{eqn:deffext}. Then we solve the system
of equations~\eqref{eqn:defsti} for the unknown
displacement vector ${\bf U}$. Its 3 last coordinates are identified as diagonal strain
components, according to definition~\eqref{eqn:defu}. 
The effective elastic properties of the packing being isotropic, we obtain $\epsilon _\alpha = \sigma_\alpha/E^*$, 
and $\epsilon _\beta = -\nu^* \sigma_\alpha/E^*$ for $\beta \ne \alpha$, in which $E^*$ and $\nu^*$
are the effective macroscopic Young modulus and Poisson
coefficient of the bead packing. On changing $\alpha$ one obtains different estimates
of those macroscopic properties, which should coincide in the limit of large systems.

This procedure is used with the numerical samples prepared in equilibrium states under varying isotropic pressure $P$, as explained in papers I and II 
(see Section~\ref{sec:assemb} below). All results are averaged over sets of 5 statistically similar samples of $n=4000$ grains each, and error bars on curves 
(which are often as small as the symbols) correspond to one standard deviation on each side of the mean value.
\subsubsection{Minimization properties}
In order to write down bounds on macroscopic elastic moduli, the following \emph{minimization properties} are useful~\cite{KR02}.
First, solving~\eqref{eqn:defsti} for ${\bf U}$ is equivalent to minimizing the following potential energy:
\be
W_1({\bf U}) = \frac{1}{2} {\bf U} \cdot \sti \cdot {\bf U} - \Delta \fext \cdot {\bf U}.
\label{eqn:min1}
\ee
Then, the contact force vector increment $\Delta {\bf f}$ minimizes
\be
W_2(\Delta {\bf f}) = \frac{1}{2}\Delta {\bf f} \cdot \kk ^{-1} \Delta {\bf f},
\label{eqn:min2}
\ee
subject to the constraint that it should be statically admissible with load increment $\Delta \fext$. Minimal values in
\eqref{eqn:min1} and  \eqref{eqn:min2} are opposite to each other, and are 
identified in the limit of large systems with the
corresponding macroscopic elastic energy, \emph{i.e.} 
$$\mp   \frac{\Omega}{2} \ww{\epsilon} : \ww{\ww{C}} : \ww{\epsilon} =  
\mp   \frac{\Omega}{2} \ww{\Delta \sigma} : \ww{\ww{C}}^{-1} : \ww{\Delta \sigma},$$
in which $ \ww{\ww{C}}$ denotes the 4th rank tensor of elastic moduli.
Those variational properties are analogous to classical results in elasticity of heterogeneous continua~\cite{NNH93}, and
will be used in Section~\ref{sec:elastp} (technical details being provided in Appendix~\ref{sec:appvar}).
\section{Properties of equilibrium configurations\label{sec:assemb}}
We summarize here some necessary information about configurations assembled by different methods, as reported in paper I~\cite{iviso1}, and then
isotropically compressed to various level of pressure, as described in paper II~\cite{iviso2}. Useful notations and properties are also briefly recalled.  
\subsection{Sample preparation and compression}
Four different configuration series which, as in papers I and II, we keep referring to as A to D, 
were prepared under a rather low pressure ($\kappa=39000$, corresponding to
10~kPa for glass beads, or $\kappa=181000$, corresponding to 1~kPa), and then quasistatically compressed up to 100~MPa 
($\kappa\simeq 80$), with friction coefficient
$\mu=0.3$ in the contacts.

They are characterized in terms of solid fraction $\Phi$, coordination number $z^*$ of the 
force-carrying structure or \emph{backbone} of the packing, proportion of \emph{rattlers} 
$x_0$ (those grains do not participate in force transmission at equilibrium), normal force distribution, friction mobilization, and geometric data such as 
distribution of interneighbor gaps and local order parameters. $z^*$ relates to the global coordination number $z=2N_c/n$ as $z^*=z/(1-x_0)$. 

Configurations A, B and D were assembled on compressing a granular gas. 
Configurations C are obtained from A and are supposed to mimic, in a simplified way, the dense 
states obtained by vibration. 

Type A samples are assembled without friction, and correspond to the ideal ``random close packing'' state (RCP), 
which according to the available numerical evidence is uniquely
defined, provided the compaction process is fast enough to avoid all incipient crystalline order nucleation~\cite[Section III]{iviso1}. 
Their solid fraction, accordingly,
is slightly below 0.64 at low pressure, while the coordination number is close to 6, with few rattlers ($x_0\simeq 1.5\%$). 
Type A configurations may thus be regarded as a simple model for 
grains that are perfectly lubricated in the assembling stage, but such that dry intergranular contacts have a frictional behavior ($\mu=0.3$) in quasistatic compression.
As a variant, another set of samples, which we denoted as the A0 series, was prepared on quasistatically compressing the solid samples
without friction. 

B states are similar to A ones, except that they are
assembled with a small coefficient of friction, $\mu_0=0.02$, as a crude model for imperfect lubrication in the fabrication stage. B states have a smaller solid fraction,
$\Phi\simeq 0.625$ instead of $0.637$ for $\kappa <10^{-4}$, 
and a slightly smaller coordination number:  $z^*\simeq 5.8$ instead of $6$ at low pressure. 
D states are the loosest of the four series, with $\Phi \simeq 0.593$ under 1~kPa, less contacts ($z^*\simeq 4.55$ at low $P$) and more than 10\% of
rattlers. Remarkably, the density of vibrated C states is close to the RCP value ($\Phi\simeq 0.635$), 
but their contact networks are as tenuous as D ones ($z^*\simeq 4.55$ at low $P$), with even more rattlers (13\%) at low pressure.
C configurations are thus denser than B ones, but much less coordinated. 

The evolution of $\Phi$, $z^*$ and $x_0$ in a pressure cycle up to 100~MPa, 
and then back to the initial value, are studied in paper II~\cite[Figs. 1 and 2]{iviso2}. 
While density increases with $P$, so do the coordination numbers, most notably above a few MPa, 
but upon decompressing many contacts are lost and coordination numbers,
if initially high, as in the A and B cases, end up with much lower values, similar to those of poorly coordinated C and D states. 
Solid fraction $\Phi$ displays very 
little hysteresis in such pressure cycles. A0 (frictionless) states behave very similarly to A ones in compression, but do not lose
their high coordination number on reducing the pressure.
The reader can refer to papers I and II for more details (\emph{e.g.}, information on
force distribution and friction mobilization). 
\subsection{Moments of normal force distributions}
In paper I, reduced moments of the distribution of normal forces $N$ were introduced, with the following notation:
\be
Z(\alpha) = \frac{\ave{N^\alpha}}{\ave{N}^\alpha}.
\label{eqn:defza}
\ee
Another quantity, closely related to $Z(5/3)$ is useful to evaluate elastic energies from contact forces. 
If $r_{TN}$ is the
ratio ${\disty \frac{\norm{T}}{N}}$ in any contact, then we define 
\be 
\tilde Z(5/3) = \frac{\langle N^{5/3}(1+\frac{5r_{TN}^2}{6\alpha_T})
\rangle }{\langle N\rangle ^{5/3}}.
\label{eqn:tilz}
\ee
$\alpha_T$ denotes the stiffness ratio defined in~\eqref{eqn:tang}.

It is convenient to use those definitions to write down averages of various quantities associated with the contacts
and proportional to some power of the normal force. As a useful example, let us relate the average normal stiffness of contacts to the pressure. 
This average, from~\eqref{eqn:kn}, reads 
$$
\ave{K_N} = \frac{1}{N_c} \sum _{i<j}  \frac{3^{1/3}}{2} \tilde E^{2/3} a^{1/3} N_{ij}^{1/3}.
$$
The sum, running over all contacting pairs $i,j$, is then transformed, using Eqns~\ref{eqn:relpn}, \ref{eqn:defza} (with $\alpha=1/3$) 
and expressing the contact density as
$$
\frac{N_C}{\Omega}=\frac{3z\Phi}{\pi a^3},
$$
into a relation conveniently displaying the dependence on pressure, solid fraction and coordination number:
\be
\ave{K_N} = \frac{3^{1/3}}{2}\tilde E^{2/3}Z(1/3)\frac{\pi^{1/3}aP^{1/3}}{z^{1/3}\Phi^{1/3}}.
\label{eqn:avkn}
\ee
This expression of $\ave{K_N}$ will be used on estimating elastic moduli. 
\subsection{Degree of force indeterminacy}
In paper I, we also discussed whether the degree of force indeterminacy of equilibrated packings could approach zero in frictional packings 
in the rigid, $\kappa\to+\infty$ limit -- this being a known property of frictionless systems~\cite{JNR2000}.
While the degree of force indeterminacy $\hhh$ is directly related to the backbone coordination number $z^*$ in frictionless packings, for which
\be
\hhh = \frac{1}{2}n(1-x_0)(z^*-6),
\label{eqn:hypersf} 
\ee
its value is more exactly evaluated, for non-vanishing intergranular friction coefficients, 
on defining a slightly corrected value of $z^*$, denoted as $z^{**}$: 
\be
z^{**} = z^*+\frac{2x_2}{3(1-x_0)},
\label{eqn:defzhhh} 
\ee
where $x_2$ is the proportion of 2-coordinated grains (which are involved in the mechanism motions mentioned in Sec.~\ref{sec:defelastic} 
and  Appendix~\ref{sec:appendixTrot}). 
Then the degree of force indeterminacy is given by 
\be
\hhh = \frac{3}{2}n(1-x_0)(z^{**}-4).
\label{eqn:hyperf} 
\ee
$x_2$ values raise to about $2.5$\% in configurations C and D, in which the ratio of the degree of force indeterminacy to the number of backbone degrees of freedom,
\emph{i.e.}, $\hhh/[6n(1-x_0)]$, does not decrease below 14\%. Only in packings assembled with an infinite friction coefficient~\cite{ZhMa05,iviso1},
called $Z$ configurations in paper I, did we obtain nearly vanishing $\hhh$ values ($\hhh/[6n(1-x_0)]$ decreasing to about 3/100).  
\section{Elastic moduli\label{sec:elastp}}
\subsection{Numerical results\label{sec:elastpnum}}
\begin{figure}[!htb]
\subfigure[$B$ versus $P$]{
\includegraphics*[angle=270,width=8.5cm]{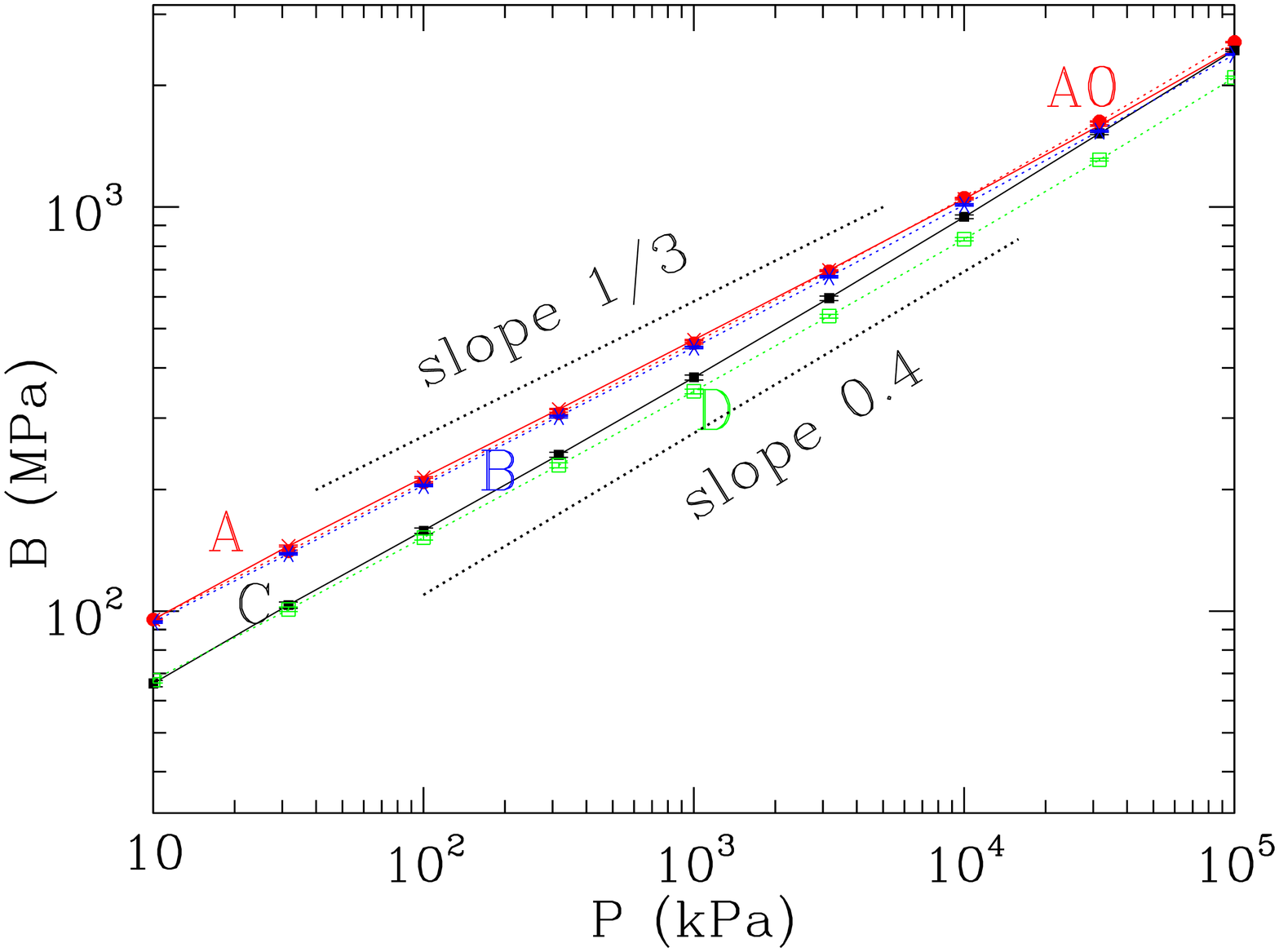}
\label{fig:modbp}}
\subfigure[$G$ versus $P$]{
\includegraphics*[angle=270,width=8.5cm]{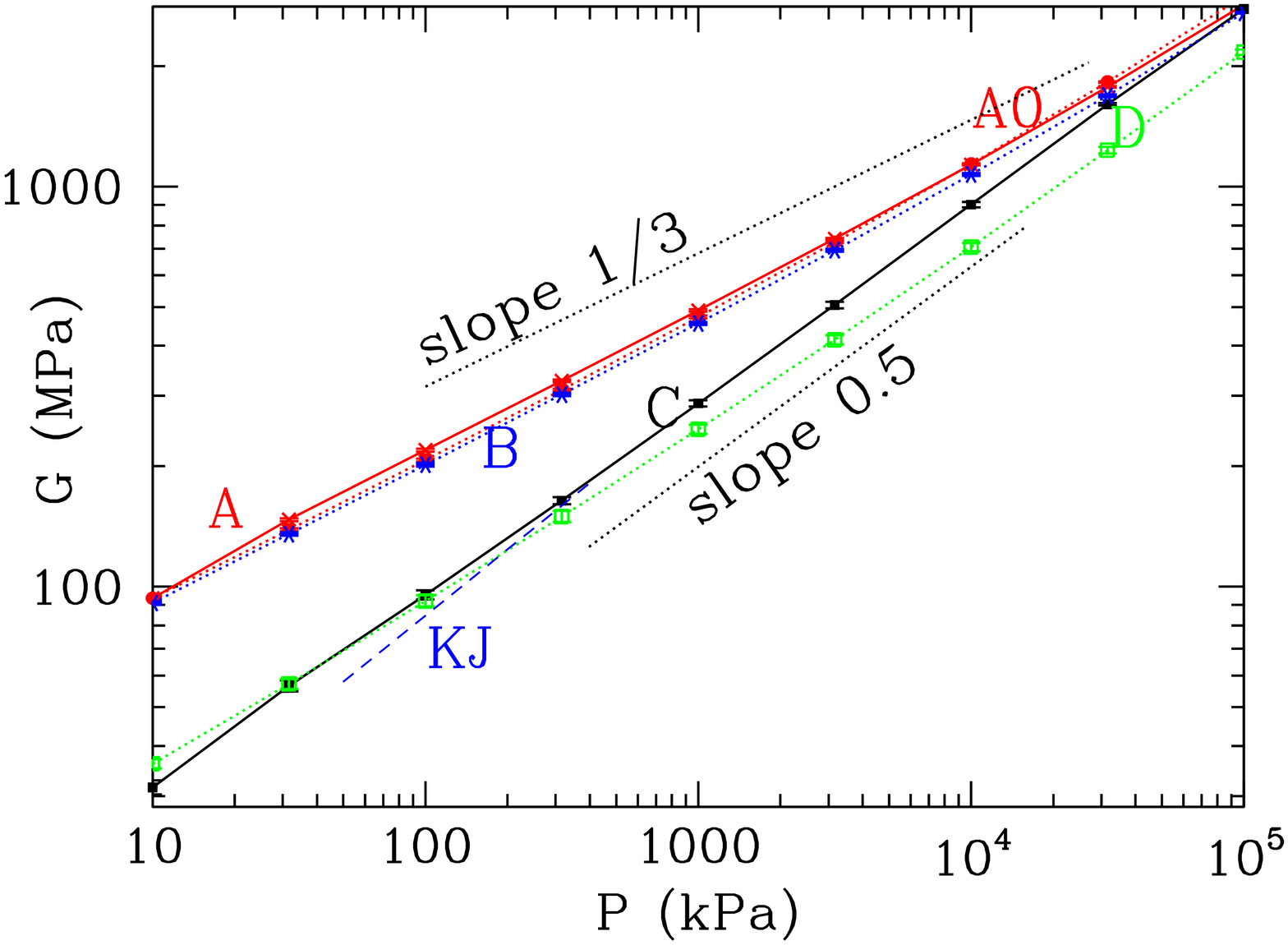}
\label{fig:modgp}}
\caption{\label{fig:modbgp}
(Color online) Bulk modulus $B$ (a) and shear modulus $G$ (b) versus confining pressure $P$ 
for series A (red, crosses, continuous line), A0 (red, round dots, dotted line),
B (blue, asterisks, dotted line), C (black, square dots, continuous line) and D 
(green, open squares, dotted line). Note that results for A, A0 and B are hardly distinguishable.
The dashed blue line marked ``KJ'' corresponds to some experimental data~\cite{KJ02} between 50 and 400~kPa commented in Section~\ref{sec:elastexp}}
\end{figure}

Elastic moduli of equilibrated configurations are evaluated as indicated in Section~\ref{sec:calcul}. Each data point on the graphs, throughout the sequel,
is based on several macroscopically equivalent load vectors in all 5 available samples for each one of the investigated macroscopic states.
Fig.~\ref{fig:modbgp} displays on logarithmic plots the pressure dependence of shear and bulk moduli in all series 
A, A0, B, C and D during the first compression.
Fig.~\ref{fig:modbgp} clearly shows that the moduli are primarily sensitive to coordination number, with well coordinated samples A, B (and A0) displaying larger moduli
than C and D, in which the contact network is more tenuous ($z^*\simeq 4.5$ and $z\simeq 4$ under low pressure~\cite{iviso1,iviso2}). Moduli are much less sensitive to 
packing fraction $\Phi$: C and D results are close to each other at low pressure, when $\Phi_C \simeq 0.638$ and $\Phi_D\simeq 0.594$~\cite{iviso1,iviso2}. 
They are not strongly influenced either by
the width of the force distribution: A and A0 states have almost the
same moduli (only some values of $G$ below 100~kPa differ by more than 5\%), whereas the probability distribution function of normal forces strongly differ as pressure grows 
(see paper II~\cite[Fig. 5]{iviso2}).

The increase of elastic constants with pressure naturally stems from the dependence of
contact stiffnesses on the force they transmit, as expressed by Eqns.~\eqref{eqn:kn} and \eqref{eqn:tang}, 
and due to relation~\eqref{eqn:relpn} the typical contact stiffness grows as
$P^{1/3}$ (see Eqn.~\ref{eqn:avkn}), which is the expected pressure dependence for macroscopic moduli. Power laws are often used 
to relate elastic moduli to confining stresses~\cite{GO90,HI96,jia99,JM01}, and
possible origins for the observation of exponents larger than $1/3$ (as on Fig.~\ref{fig:modbgp}) have been discussed by several authors~\cite{GO90,dG96}. 
One possible explanation is the creation of new
contacts under the effect of the increase of the confining pressure, which leads to a denser, stiffer contact network. 
This mechanism appears in particular to account for
the pressure dependence of elastic moduli in regular, crystal-like arrays of identical particles as in the experiments 
described in Refs.~\cite{DM57} and~\cite{gilles03}.
Because of the slight lattice distorsions obtained with imperfect and slightly polydisperse spheres, 
the contact coordination number, which is limited, in the rigid limit of
$\kappa\to \infty$, to 4 in 2D and 
6 in 3D~\cite{JNR2000}, is smaller than the nearest neighbor coordination number on the dense lattices studied 
(such as 12 for FCC in 3D \cite{DM57} and 6 for the 2D triangular
lattice~\cite{gilles03}). This leaves a large number of neighbor pairs at a distance related 
to the width of the particle size distribution, where additional contacts are induced by
higher pressures. This has been shown by numerical simulations~\cite{JNR97a} to produce a pressure 
dependence of moduli closer to $P^{1/2}$ in some pressure range, a phenomenon predicted
in part by a theory presented in~\cite{velicky02}. With general, amorphous packings, 
the situation is different because distances between neighbors that are not in contact are no longer related
to a small polydispersity parameter, but are distributed, approximately as a power law in some range (see paper I~\cite{iviso1}),
in a way that is characteristic of the disordered geometry. Departures from the $P^{1/3}$ scaling are larger in low $z$ states (Fig.~\ref{fig:modbgp}), 
and the largest in C configurations, in which contact gains under growing $P$ are faster than in D ones.
However, apparent power laws with exponents larger 
than $1/3$ are observed at very low pressures, when, from paper II~\cite[Fig.2a]{iviso2},
%Fig.~\ref{fig:zp},
the increase of $z$ with $P$ is rather slow. Moreover, in the case of C and D systems, 
the exponent of the power law fit for the pressure dependence of shear modulus $G$ is 
significantly larger (about $0.5$) than the one for bulk modulus $B$ (about $0.4$). 
These features are discussed in paragraph~\ref{sec:elastpred} below. Changes of ratio $G/B$ as $P$
grows are equivalent to changes of the Poisson ratio of the granular material, given by
\be
\nu^* = \frac{3B-2G}{6B+2G}.
\label{eqn:defnu}
\ee
%The evolution of the Poisson ratio with growing pressure is shown on Fig.~\ref{fig:nup} for the different sample series. 
$\nu^*$ decreases only 
slightly as $P$ grows for well coordinated states A and B, from $\nu^*\simeq 0.13$ 
at P=10~kPa to $\nu^*\simeq 0.09$ under 100~MPa. Its larger variations in poorly coordinated configurations C and D,
for which it decreases from 0.3 to about 0.1 in the same range, corresponds to $G$ increasing with $P$ faster
than $B$.
\subsection{Simple prediction schemes and relations to microstructure\label{sec:elastpred}}
\begin{figure}[htb]
\subfigure[$b_r$ versus $(z\Phi)^{2/3}$]{
\includegraphics*[angle=270,width=8.5cm]{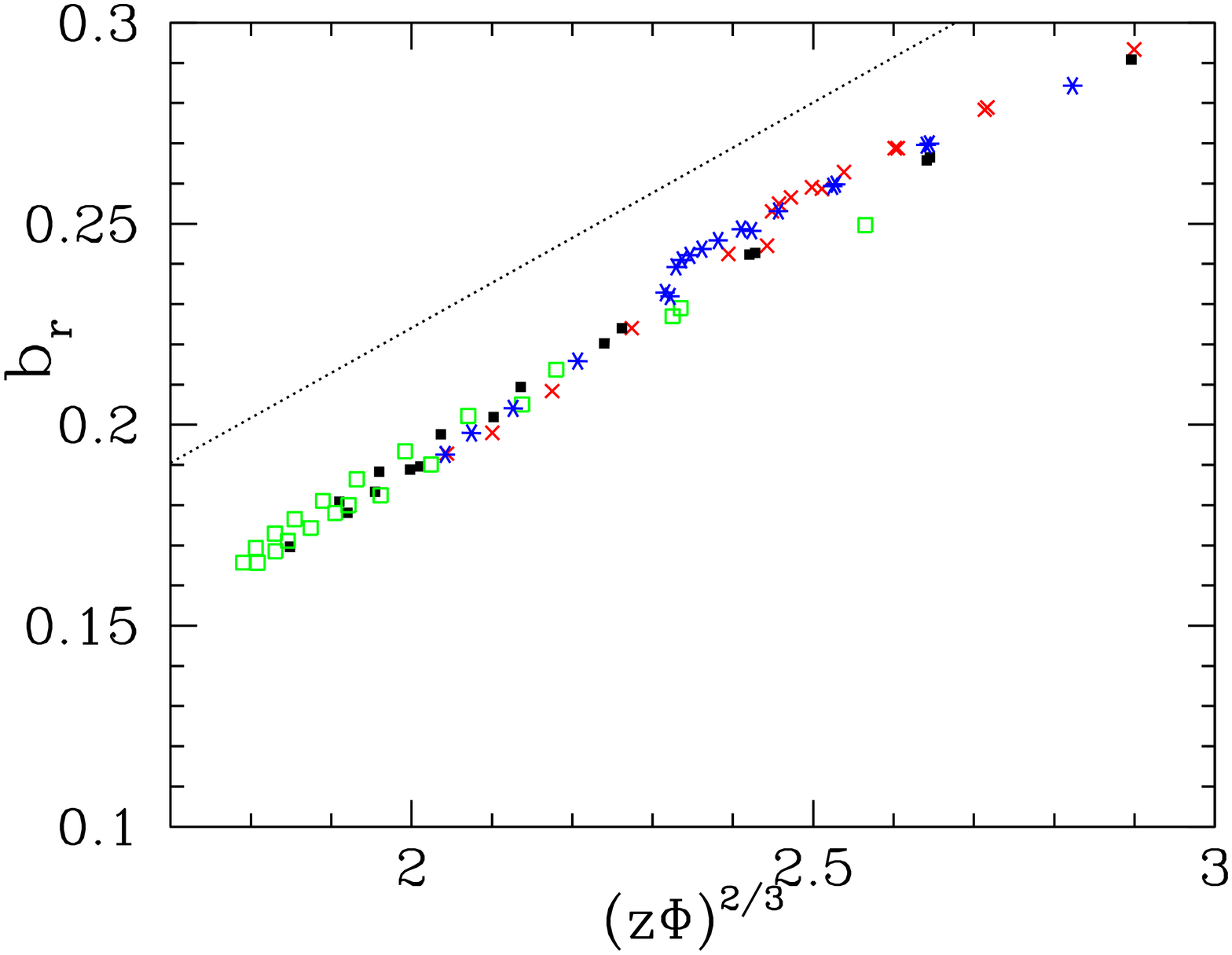}
\label{fig:modbtout}}
\subfigure[$g_r$ versus $(z\Phi)^{2/3}$]{
\includegraphics*[angle=270,width=8.5cm]{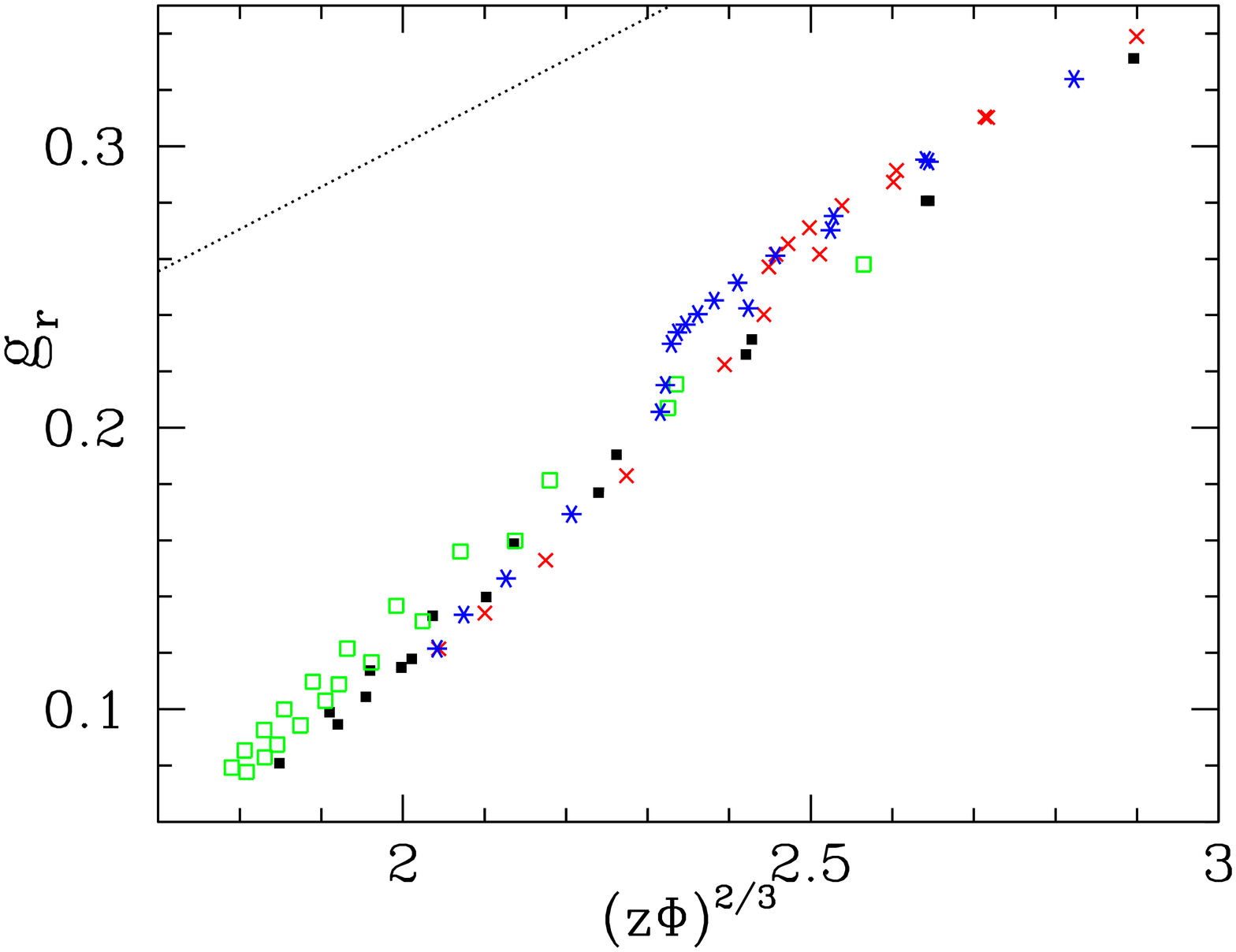}
\label{fig:modgtout}}
\caption{\label{fig:modbgtout}
(Color online) Reduced moduli, as defined in~\eqref{eqn:brgr}, in units of $\tilde E ^{2/3}P^{1/3}$ as functions of 
$(z\Phi)^{2/3}$, same colors and symbols as on Fig.~\ref{fig:modbgp}.
The estimates given in~\eqref{eqn:bge} are plotted as straight dotted lines. 
Moduli are plotted for all configurations in the pressure cycle, showing an approximate collapse on 
a single curve.}
\end{figure}

The simplest approximation scheme to estimate the values of elastic moduli, knowing the density and the coordination number, is based
on the assumption of homogeneous strains (or, equivalently, of affine displacements). It was introduced, \emph{e.g.} in \cite{WAK87}, and it is
also used by Makse \emph{et al.} in Refs.~\cite{MGJS99,Makse04} (where it is called an effective medium theory). 
It amounts to evaluating the stress increments corresponding to strain $\ww{\epsilon}$
using formula~\eqref{eqn:stress}, in which the contact force variations are evaluated, \emph{via} Eqn.~\eqref{eqn:cmat}, with relative displacements given by
$$
{\bf u}_i - {\bf u}_j = \ww{\epsilon}·\cdot ({\bf r}_j - {\bf r}_i).
$$
Using the isotropy of the distribution of contact orientations, and replacing all normal forces by their average value 
in the computation of contact stiffnesses, this results,
using relation~\eqref{eqn:relpn}, in the following estimates.
\be
\ba
B^{\text{e}}=\frac{1}{2}\left(\frac{z\Phi\tilde E}{3\pi}\right)^{2/3}P^{1/3}\\
G^{\text{e}}=\frac{6+9\alpha_T}{10} B^{\text{e}}.
\ea
\label{eqn:bge}
\ee
One thus finds the expected $P^{1/3}$ dependence, and obtains moduli proportional to $(z\Phi)^{2/3}$. 
Formulae~\eqref{eqn:bge} also predict a constant $G/B$ ratio, and thus
a constant Poisson ratio:
\be 
\nu^e= \frac{6(1-\alpha_T)}{26+9\alpha_T}\simeq 0.032
\label{eqn:nue}
\ee
This latter estimation is considerably smaller than the measured values which are given above (shortly after Eqn.~\ref{eqn:defnu}),
as noted in~\cite{Makse04}. This mainly stems from the inaccuracy of the estimated value of $G$~\cite{MGJS99},
as we shall see. Eqn.~\ref{eqn:bge} suggests to represent ratios 
\be
\ba
b_r &= \frac{B}{\tilde E ^{2/3}P^{1/3}}\\
g_r &= \frac{G}{\tilde E ^{2/3}P^{1/3}}
\ea
\label{eqn:brgr}
\ee 
as functions of $(z\Phi)^{2/3}$, which is done on Fig.~\ref{fig:modbgtout}.
Fig.~\ref{fig:modbgtout} shows that $G^e$ is a significantly poorer estimate of $G$ than $B^e$ of $B$, as already noted in~\cite{MGJS99}, for samples of type A or B, 
and even more so in low coordination number configurations C and D. It also shows that the elastic moduli, as a first approximation, 
can be thought of as determined by
$z$ and $\Phi$, the former quantity, as it varies more between different sphere packings, being the most influential. 
The present study can thus be regarded as a first step towards
a method to infer coordination numbers nfrom the elastic moduli. As stressed in paper I,
coordination numbers are virtually inaccessible to direct measurements.
(It should nevertheless be recalled that the present work is limited
to \emph{isotropic} configurations, implying isotropy of fabric as well as isotropy of stresses). 
In this respect it is interesting to note that the configurations of lower coordination number obtained upon
\emph{decompressing} A and B ones after they first reach a high pressure level (see paper II~\cite{iviso2}) yield data points on Fig.~\ref{fig:modbgtout} that stay
close to the C and D ones corresponding to the same product $z\Phi$.

\begin{figure}[htb]
\includegraphics*[angle=270,width=8.5cm]{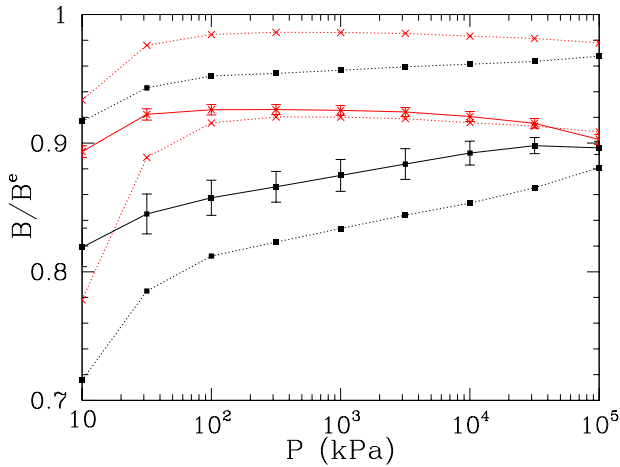}
\caption{\label{fig:bappr}
(Color online) Ratio $B/B^e$ (symbols connected with continuous line, error bars), 
and its Voigt and Reuss bounds (symbols connected by dotted lines) in configurations 
A (red, crosses) and C (black, square dots), during compression. B and D samples respectively behave similarly to A and C.
}
\end{figure}

An interesting alternative to the direct use of formula~\eqref{eqn:stress} is to exploit the variational property expressed by~\eqref{eqn:min1}, as explained
in Appendix~\ref{sec:appvar}. This shows that $B^e$ and $G^e$ are upper bounds to the true moduli. 
Accounting for the distribution of forces (see the derivation of Eqn.~\ref{eqn:avkn}), those bounds
can be slightly improved, yielding the analogs of the Voigt upper bound for the macroscopic elastic 
moduli of a mesoscopically disordered continuous material:
\be
\ba
B&\le B^{\text{Voigt}}=B^eZ(1/3)\\
G&\le G^{\text{Voigt}}= G^eZ(1/3),
\ea
\label{eqn:bgvoigt}
\ee
where $Z(1/3)$, as defined in~\eqref{eqn:defza}, is always strictly smaller than 1. 
For the bulk modulus, one can also take advantage of the second variational
property, expressed by~\eqref{eqn:min2}. As explained in Appendix~\ref{sec:appvar}, 
this requires a trial set of contact force increments which balance the applied
load increment. When the stress increment is proportional to the preexisting stress, 
one may take increments of contact forces that are also proportional to their initial values. No such
forces balancing a shear stress are available for isotropically prestressed configurations. 
One thus obtains (see Appendix~\ref{sec:appvar} for details) a \emph{lower} bound for
$B$ which is analogous to the Reuss bound for the macroscopic elastic bulk modulus of a mesoscopically disordered continuous material. 
This lower bound $B^{\text{Reuss}}$
involves the dimensionless quantity $\tilde Z(5/3)$ defined in~\eqref{eqn:tilz}, and enables one to bracket the bulk modulus:
\be
\frac{B^e}{\tilde Z(5/3)}=B^{\text{Reuss}}  \le B \le  B^{\text{Voigt}}=B^eZ(1/3).
\label{eqn:encB}
\ee
The ratio of the upper bound to the lower one is therefore related to the shape of the distributions of normal forces and to the mobilization of friction.
Fig.~\ref{fig:bappr} displays ratios $B/B^e$, $B^{\text{Voigt}}/B^e$, and $B^{\text{Reuss}}/B^e$ versus (growing) pressure $P$ in configurations A and C.
These data show that in both cases of high (A) and low (C) initial coordination number, the bracketing of $B$ given by~\eqref{eqn:encB} is quite
accurate, the relative difference
between upper and lower bounds staying below 10\% except at the lowest pressure for A, and around 15\% for C. 
The Reuss estimate is better than the Voigt one in general. It is
even excellent in the A case for all but the two lowest pressure values studied 
($B$ appears to be slightly smaller than its lower bound at very high pressure because we neglected 
the reduction of intercenter distances between contacting grains in the evaluation of the bounds). 
This estimate (see Appendix~\ref{sec:appvar}) becomes exact when the trial force increments used are the right ones, and assumes therefore that the
shape of the normal force distribution and  the level of friction mobilization
do not change with pressure. The observation, reported in paper II~\cite{iviso2}, that 
such quantities are indeed nearly independent on pressure in A states (except at the lowest pressure), is therefore consistent with the
success of the Reuss type estimate for $B$ in that case.

Yet, for $G$ no Reuss estimate is available, and the use of the Voigt one with the factor $Z(1/3)$ hardly reduces the discrepancy between $G^e$ and $G$:
this factor can be read on Fig.~\ref{fig:bappr}, where it coincides with the upper bound, and hence
$G^{\text{Voigt}}$ is only smaller than $G^e$ by a few percent. It thus overestimates the true 
shear modulus by 30 to 40\% in well-coordinated states, and even by a factor of 3 in
poorly coordinated ones at low pressure. 
\subsection{Fluctuations and more sophisticated prediction schemes.\label{sec:sophi}}
\begin{figure}[htb]
\includegraphics*[angle=270,width=8.8cm]{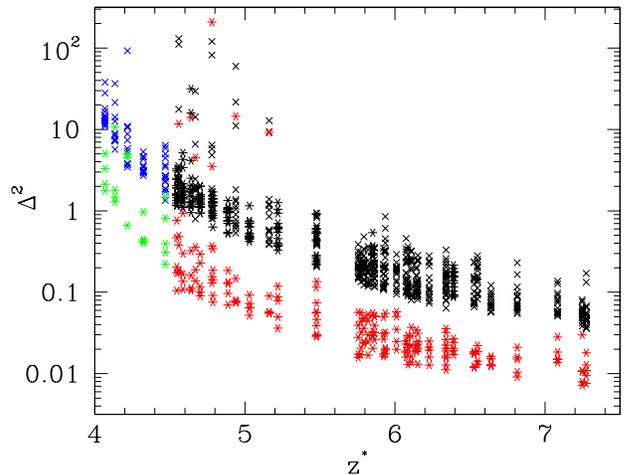}
\caption{\label{fig:dfluc}
(Color online) $\Delta ^2$, as defined in~\eqref{eqn:delta2}, versus backbone coordination number $z^*$, 
for isotropic stress increments (red asterisks, green for Z states) and
pure deviatoric ones (black crosses, blue for Z states).
}
\end{figure}

The Voigt or mean field approach ignores fluctuations in grain displacements and rotations. 
As an indicator of the amplitude of such fluctuations is the average of squared
particle displacements, we measured the ratio
\be
\Delta ^2 = \frac{1}{n^*\norm{\ww{\epsilon}}^2}\sum _{i=1}^{n^*} \vert\vert \tilde {\bf u}_i  \vert\vert^2,
\label{eqn:delta2}
\ee
in which the squared strain amplitude, $\norm{\ww{\epsilon}}^2=\epsilon_1^2 +\epsilon_2^2 +\epsilon_3^2$, 
normalizes the average of squared displacement fluctuations, as defined in~\eqref{eqn:utilde}, the sum running over the $n^*=n(1-x_0)$ force-carrying grains. 
Fig.~\ref{fig:dfluc} displays the values of $\Delta^2$ evaluated in all samples at the different values of the confining pressure. 
$\Delta^2$ is distributed over some fairly wide
interval in similar configurations, but is systematically larger for purely deviatoric stress increments than for isotropic pressure steps and
has a clearcut decreasing trend as a function of backbone coordination number $z^*$.
(To add data points with lower $z^*$ on Fig.~\ref{fig:dfluc} we also used 
the Z series, infinite friction samples, as described in paper I and Section~\ref{sec:assemb}, isotropically compressed and equilibrated at five pressure levels,
from 1 to 100~kPa).
This is consistent with the approximation that ignores fluctuations being less accurate for shear stresses than for isotropic ones.

More elaborate prediction methods for elastic moduli were proposed. Kruyt and Rothenburg~\cite{KR02,KR04} considered two-dimensional assemblies of non-rotating
particles, and succeeded in applying a variational approach such as the one we use for bulk modulus $B$ to the evaluation of shear moduli as well. 
Velick\'y and Caroli~\cite{velicky02} studied the case of an imperfect lattice system with contact disorder, as in the experiments of~\cite{DM57} and the 
simulations of~\cite{JNR97a}. Jenkins \emph{et al.}~\cite{JimLuigiflu05} dealt with frictionless sphere packings. More recently, La Ragione and Jenkins~\cite{LaJe07}
published an approximation scheme which is directly comparable to our simulation results, the results of which are
denoted as LRJ below. 
%For completeness, the corresponding formulae are written down in Appendix~\ref{sec:appLRJ}.

Those estimation procedures improve upon the Voigt assumption that relative displacements are ruled by the average strain 
on considering small sets of displacements and rotations, either associated to one grain, or to a contacting pair. Those 
degrees of freedom are allowed to fluctuate while their
surroundings abide by the Voigt assumption. Optimal values of the fluctuating variables are then to be determined on solving 
the corresponding system of equilibrium
equations for the selected small set of degrees of freedom.
Such approaches necessarily 
involve complex treatments of the random geometry of local grain arrangements, especially on attempting to express the predicted moduli with a limited 
amount of statistical data. 
In~\cite{iv2} we numerically check some of the  approximations involved, on 
exactly solving the required set of local equilibrium problems. 
We show~\cite{iv2} that the discrepancy between observed and predicted shear moduli is reduced, down to 50\% in the worst cases 
of C and D samples under low pressures. 

\begin{figure}[!h] 
\centering
\includegraphics*[width=8.5cm]{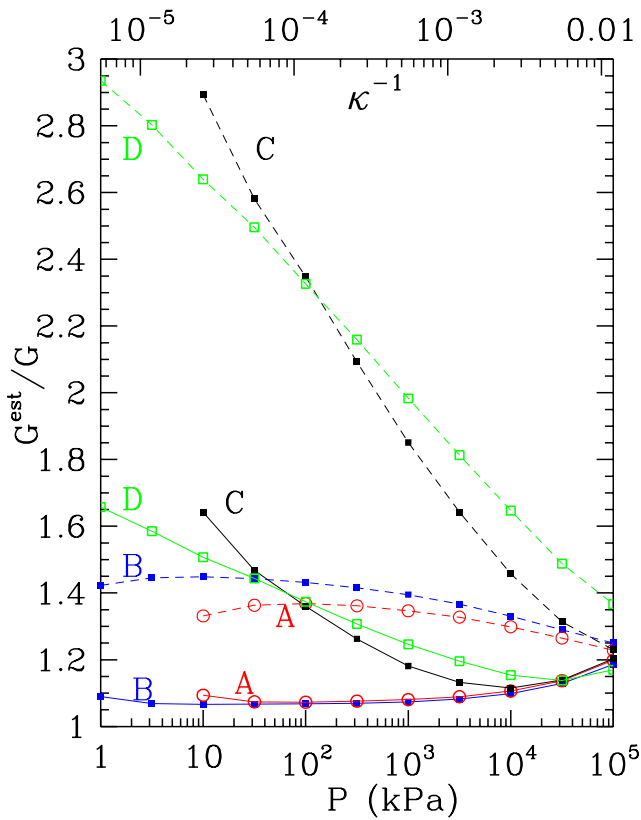}
\caption{\label{fig:luigi2} (Color online) Ratio of estimated to measured values of shear moduli, $G^{\text{est}}/G$, with $G^{\text{est}}=G^{\text{Voigt}}$ 
(larger values, dots connected by dashed lines) and $G^{\text{est}}=G^{\text{LRJ}}$ 
(smaller values, dots connected by solid lines), versus $P$ or $\kappa^{-1}$, in sample series A (red), B (blue), C (black) and D (green).
}
\end{figure}
Leaving aside the discussion of the various approximations involved in the derivation of the LRJ predictions (some important steps
of which are tested in~\cite{iv2}), we now confront them with the numerical data.
  
We  observe that the LRJ formulae do not improve the predictions of bulk moduli over the Voigt and Reuss bounds~\eqref{eqn:encB}. 
Occasionally, they even predict values \emph{below} the lower bound. 
Yet,
as shown on Fig.~\ref{fig:luigi2}, the estimates  $G^{\text{LRJ}}$ obtained for shear moduli are much better than the Voigt ones~\eqref{eqn:bgvoigt}.
Shear moduli in well-coordinated states A and B are accurately
predicted, while the discrepancy in poorly coordinated systems C and D, from a factor of three with the Voigt formula, are down to about 
50\%-70\% with the LRJ one under the lowest pressure levels.

LRJ formulae yield moduli that are proportional to average contact stiffnesses (in which we introduced an additional factor of $Z(1/3)$, see Eqn.~\ref{eqn:avkn}), 
with coefficients involving  rational functions of the backbone coordination number $z^*$, 
and also the variance of the fluctuations in the number of contacts of backbone grains.
The LRJ predictions still overestimates the shear modulus in poorly coordinated systems, for which we now test another kind of theoretical prediction. 
\section{The case of poorly coordinated networks\label{sec:elast0}}
The specific elastic properties of configuration series C and D, with their small coordination numbers, 
are reminiscent of frictionless packings~\cite{OSLN03,Makse04},
in which a similar anomalous behavior of $G$ as a function of pressure has been reported. Here we review these properties of packings with no tangential
forces (Section~\ref{sec:elast0sf}), and we 
discuss a possible explanation~\cite{Wyart-th}, and its applicability to poorly coordinated frictional packings (Section~\ref{sec:elast0f}). 
\subsection{Frictionless packings\label{sec:elast0sf}}
\begin{figure}[htb]
\subfigure[Bulk modulus.]{
\includegraphics*[angle=270,width=8.5cm]{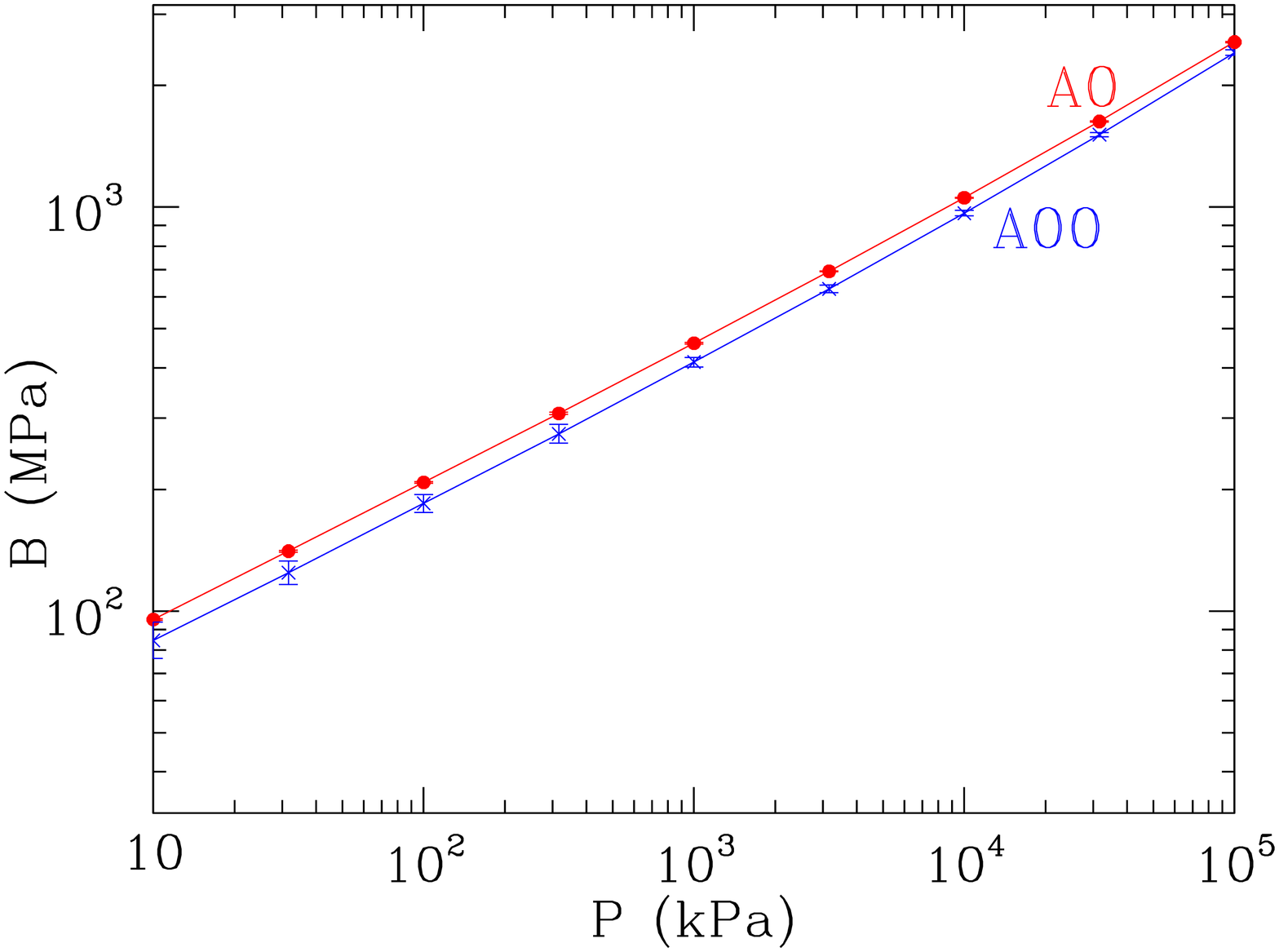}
\label{fig:modbsf}}
\subfigure[Shear modulus.]{
\includegraphics*[angle=270,width=8.5cm]{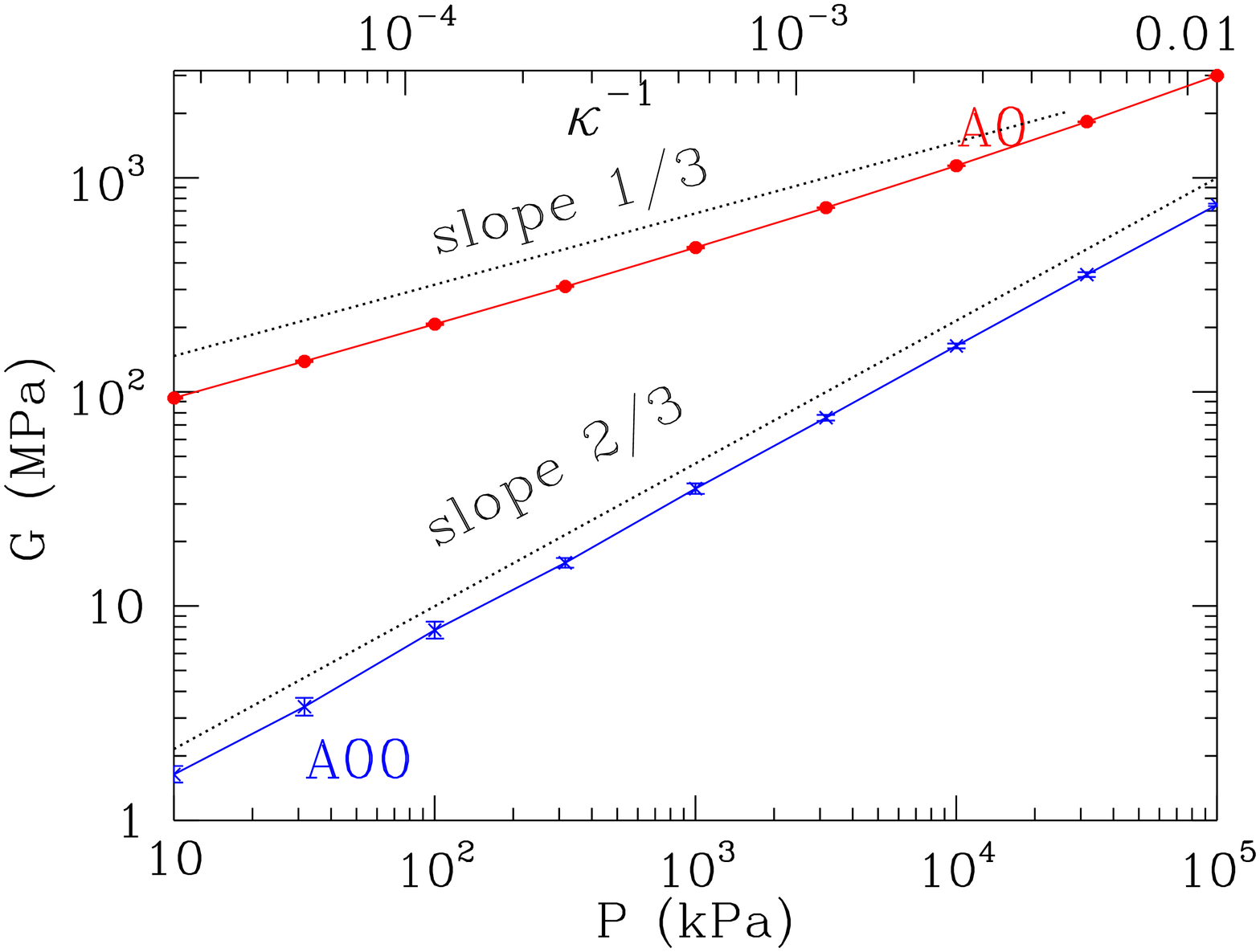}
\label{fig:modgsf}}
\caption{\label{fig:modsf}
(Color online)
Elastic moduli of samples A0 prepared at different pressures without friction, computed with (A0) and 
without (A00) tangential elasticity, A00 results corresponding to completely frictionless packings.}
\end{figure}

Although samples of series A0 were confined with no mobilization of friction, elastic moduli shown on Fig.~\ref{fig:modbgp} 
have been computed with tangential elasticity in
the contacts, just like, \emph{e.g.} in  Ref.~\cite{MGJS99}. It is assumed for state A0 
that friction is not mobilized in the preparation process. In other words samples are perfectly annealed to a local minimum of 
mechanical energy inconfiguration space, but the response to some stress increment implies tangential forces in the contacts. 
Results are of course different if contacts are still regarded
as frictionless on evaluating elastic properties. Fig.~\ref{fig:modsf} compares this new set of values, which we denote as A00, to A0 ones.  
Bulk moduli (Fig.~\ref{fig:modbsf}) are
only slightly higher (about 10$\%$ at low pressure) with tangential elasticity.  A00 values correspond to frictionless packings, in which
a relation between pressure $P$ and the increase in solid fraction $\Delta \Phi$ above the rigidity threshold (\emph{i.e.}, above the value of $\Phi$ in the
limit of vanishing pressure) was obtained in paper I~\cite[Eqn. 31]{iviso1}, as a direct consequence of the isostaticity property of the backbone.  
It is straightforward to check that this relation, on taking the derivative with respect to $\Phi$ and relating modulus $B$ to $P$, 
yields the Reuss type estimate of $B$, as expressed by~\eqref{eqn:encB} and \eqref{eqn:bge}, in which 
$z$ and $\Phi$ are replaced by their values in the limit of $P\to 0$ (which does not significantly affect the result at low pressure). 
Both approaches are based on the assumption that forces vary proportionnaly to $P$, which is a particular consequence of isostaticity, hence the exactness of the Reuss
estimate for A00 results. The distribution of 
force values in sample series A0, once 
normalized by the applied pressure level, was checked in paper II~\cite{iviso2} to remain very nearly constant.

The small influence
of tangential elasticity on bulk moduli, which is responsible for the difference in $B$ values between A0 and A00 series, is not surprising, as both the 
Voigt and the Reuss-like approaches, which restrict the values of $B$ to the interval given by~\eqref{eqn:encB}, 
lead to the assumption of vanishing tangential forces. 

\begin{figure}
\centering
\includegraphics*[angle=270,width=8.5cm]{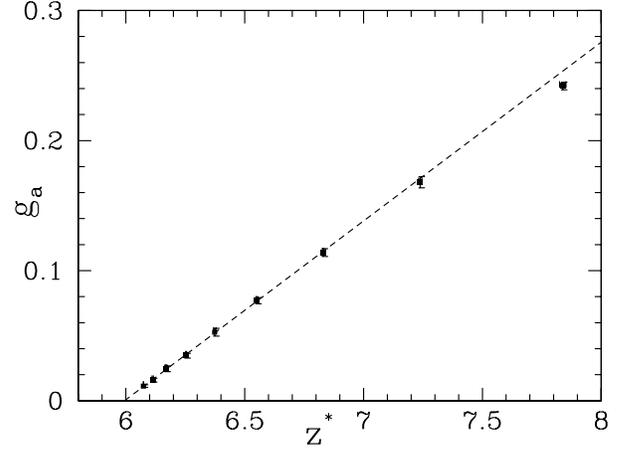}
\caption{\label{fig:gz0} Reduced shear modulus $g_a$, defined by~\eqref{eqn:defga}, in frictionless configurations A00 versus backbone coordination number $z^*$.
The straight line is the best linear fit through the 6 leftmost data points.}
\end{figure}

The shear modulus, on the other hand, as noted in refs.~\cite{OSLN03,Makse04}, is singular in frictionless packings under isotropic stresses: 
while values of $G$ in the A0 series vary approximately proportionally to $B$, and are of the same order, as observed above in Sec.~\ref{sec:elastpnum}, 
shear moduli of frictionless systems
A00 (Fig.~\ref{fig:modgsf}) are considerably smaller, and vary faster with $P$. This increase is very well fitted
by a power law with exponent $2/3$, in agreement with~\cite{OSLN03}. 
An explanation for the singular behavior of $G$ is suggested in~\cite{Wyart-th,Wyart-th2}, as follows. 
First, some of the pressure dependence of $G$ is simply
due to the influence of pressure on average contact stiffness $\ave{K_N}$.
One should therefore rather explain the pressure
dependence of $G/\ave{K_N}$. Then it is argued that this amplitude is proportional to the degree of force indeterminacy, or to
$z^*-6$. More precisely, 
the shear modulus should scale as the degree of force indeterminacy per unit volume, or equivalently as $(z^*-6)(1-x_0)\Phi$. 
This is the crucial part of the argument. Leaving aside a discussion of its justification (which would require detailed calculations of
sets of self-balanced contact forces and response functions within the contact networks) we check here for its practical validity. 
To do so, we define a reduced shear modulus $g_a$ on dividing by $(1-x_0)\Phi$ and by $\ave{K_N}$. We thus have, from~\eqref{eqn:avkn},
\be
g_a=\frac{Gz^{1/3}}{\tilde E^{2/3}P^{1/3}Z(1/3)(1-x_0)\Phi^{2/3}},
\label{eqn:defga}
\ee
and we test (Fig.~\ref{fig:gz0}) whether it varies linearly with $z^*$.

The final part of the demonstration of Ref.~\cite{Wyart-th} (see also~\cite{ZhMa05}), 
suggests to evaluate the increase of coordination number with pressure on relating both quantities to
the increase in packing fraction above rigidity threshold $\Delta \Phi$. Such a relation between $P$ and 
$\Delta\Phi$ in isostatic frictionless packings was written
in~\cite{iviso1}, with $P\propto (\Delta\Phi)^{3/2}$.  
The additional ingredient is a scaling form of the
increment of $z^*$ with $\Delta \Phi$:
\be
z^*-6 \propto \Delta \Phi ^{1/2}.
\label{eqn:dzdphi}
\ee
A homogeneous shrinking approximation is suggested in~\cite{Wyart-th} to derive relation~\eqref{eqn:dzdphi}, based on 
the assumption that the gap-dependent near neighbor coordination number $z(h)$ grows like $z(h)-6 \propto h^{0.5}$. However, as shown in~\cite{iviso1} 
we observed an exponent $0.6$ instead, and our data therefore do not confirm this argument.

Nevertheless, the proportionality of the singular amplitude $g_a$ of the shear modulus to $z^*-6$ is accurately satisfied, as shown on Fig.~\ref{fig:gz0}.
The linear fit of the dependence of $g_a$ on $z^*$, through the 6 first data points, is very good and predicts a vanishing modulus for $z^*=5.994\pm 0.008$.
\subsection{Packings with intergranular friction \label{sec:elast0f}}
\begin{figure}[!htb]
\includegraphics*[angle=270,width=8.5cm]{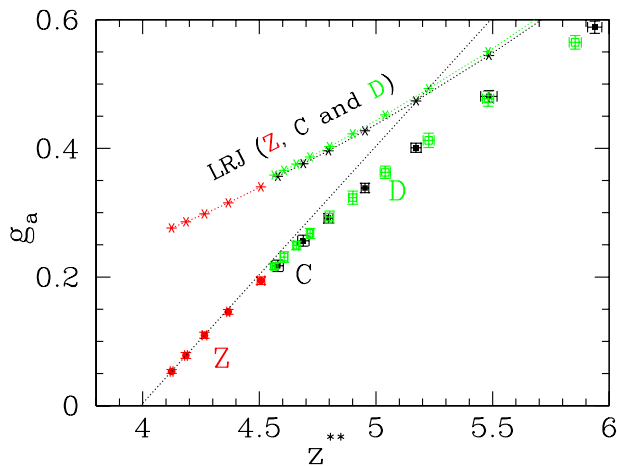}
\caption{\label{fig:gzQEZ} (Color online)
Amplitude $g_a$ (Eqn.~\ref{eqn:defga}) versus $z^{**}$, as defined in \eqref{eqn:hyperf}, in samples of types C (black), D (green) and Z (5 lower data points, red).
The dotted line is a linear fit to the 4 lowest Z data points. The LRJ predictions are shown for 
series z, C and D, as indicated (asterisks joined by dotted lines, same color code).}
\end{figure}

In paper I we concluded that frictional packings prepared in low coordination states did not approach isostaticity under low pressure.
However, one may test whether the amplitude $g_a$ varies linearly with the degree of force indeterminacy when it is small enough, 
even if it does not approach zero. The Z states, on the other hand (see Section~\ref{sec:assemb} and paper I~\cite{iviso1}), 
were prepared with an infinite friction coefficient
and have nearly vanishing force indeterminacy at low pressure.
Fig.~\ref{fig:gzQEZ}, in which $g_a$ data for states
C, D and Z are plotted versus the corrected backbone coordination number $z^{**}$, which determines
the degree of hyperstaticity per degree of freedom by ~\eqref{eqn:hyperf}, 
shows that the linearity is very well satisfied for Z states. Z, C and D points lie approximately on the same curve, showing that
the macroscopic modulus is controlled by $z^{**}$, in spite of the differences in the structures of states C, D and Z. 
(Due to the greater micromechanical changes observed upon \emph{reducing}
the pressure from its maximum value~\cite{iviso2}, the corresponding data points for  C and D states lie on a different curve, not shown on the figure.)
The linear fit still approximately applies to the lowest values for D and C. 
As expected, the linear fit predicts $G=0$ for $z^{**}=3.99\pm 0.02$, \emph{i.e., } a shear modulus vanishing proportionnally to the degree of force indeterminacy. 
Z configurations have a larger population of rattlers and divalent grains than C or D ones: $x_0\simeq 0.184$ and $x_2\simeq 0.068 $ at 1~kPa.
Consequently, on fitting $g_a$ versus $z^*$ instead of $z^{**}$, $g_a$ would appear to vanish unambiguously below $z^*=4$, for the
value $z^*=3.93\pm 0.02$. The correction due to 2-coordinated beads improves the accuracy of the result that $G$ vanishes with $H$.

Fig.~\ref{fig:gzQEZ} also displays the values of amplitudes $g_a$ predicted by the LRJ formulae~\cite{LaJe07}, as discussed in Section~\ref{sec:sophi}. As should
normally be expected for such an estimation procedure, based on the local equilibrium of one pair of grains embedded in an elastic medium, the LRJ approach is unable
to capture the vanishing of shear moduli in the limit of $z^{**} \to 4$ 
(\emph{i.e.}, $\hhh \to 0$), since the rigidity properties of tenuous networks are determined by more 
collective effects.  The low level of force indeterminacy provides a complementary approach to the estimation schemes evoked in Section~\ref{sec:sophi}
to predict the values of shear moduli in isotropically compressed packings.

We conclude therefore that the proximity of a state devoid of force indeterminacy, however unreachable, explains the anomalously fast increase
of the shear modulus with the pressure for low coordination frictional packings, as observed on Figs.~\ref{fig:modbgp} and \ref{fig:modbgtout}.
As to the increase of the degree of hyperstaticity, or of $z^{**}$, with pressure, its prediction seems to be even more difficult than in the frictionless case. 
What would be needed is an accurate prediction of small changes in $z^*$, which, as observed in~\cite{iviso2} (paper II) the simple
homogeneous shrinking assumption does not provide, due, in particular, but not only, 
to its inability to deal with the recruitment of rattlers by the growing backbone.

The proximity of a ``critical'' value of the number of contacts on the backbone also entails specific properties of the eigenvalues 
of the stiffness matrix (the ``density
of states'' in the language of solid-state physics), with a large excess of soft modes~\cite{OSLN03,WNW05}. 
A similar behavior, both for the eigenmodes of the stiffness matrix and
for some shear elastic constant was observed in~\cite{SRSvHvS05} in 2D simulations of anisotropic states. 
From this particular set of results in anisotropic packings and from the
Reuss approach to estimate the bulk modulus one may deduce that the non-singularity of $B$, 
as opposed to $G$, directly stems from the \emph{isotropic} state of stress on which
load increments are applied. On increasing $P$, in a good approximation (the better the lower the degree of force indeterminacy), 
one just rescales the contact force values.
Load increments that are not proportional to the preexisting load, on the other hand, tend to produce large fluctuations (see Fig.~\ref{fig:dfluc})
and soft responses in poorly coordinated contact networks.
\section{Comparison with experimental results\label{sec:elastexp}}
%\subsection{Lubricated and vibrated packings}
The assembling procedures of states B and C (Sec.~\ref{sec:assemb}) can be regarded as idealized 
models for lubrication and vibration. Jia and Mills~\cite{JM01,ARMJM05} measured sound wave velocities in glass bead packings, some 
samples being densified by repeated taps on the container, and others mixed with a very small quantity of a lubricant (trioleine).
The beads were placed in a cylindrical container and then compressed by a piston, transmitting a confining pressure. Velocities
of longitudinal and transverse sound waves propagating in the vertical direction (orthogonal to the piston) were measured
in the 70~kPa-800~kPa range. Those velocities, which we denote as usual as $V_P$ and $V_S$, relate for an isotropic material to bulk and
shear moduli and mass density $\rho$ as
$$
V_P = \sqrt{\frac{B+\frac{4}{3}G}{\rho}}\ \mbox{and}\ V_S = \sqrt{\frac{G}{\rho}}.
$$
\begin{figure*}[!htp]
\centering
\includegraphics*[width=12cm]{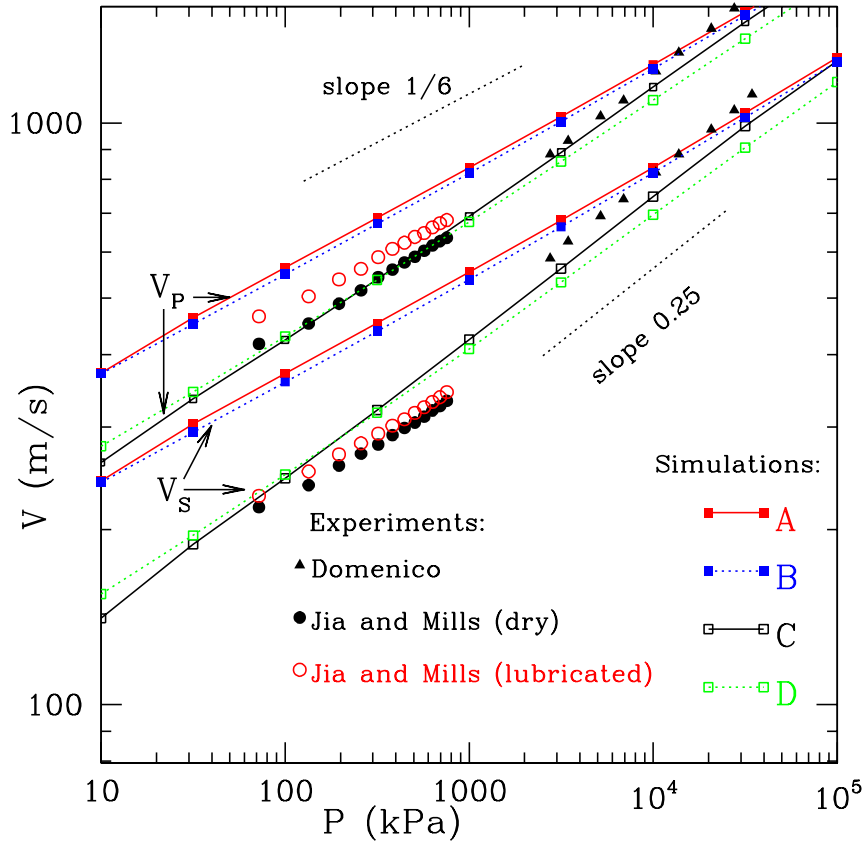}
\caption{(Color online) Sound velocities $V_P$ (upper date points) and $V_S$ (lower data points) 
versus confining pressure $P$ on double logarithmic plot for
the experimental dry (vibrated) and lubricated samples of Jia and Mills, and of Domenico, to which numerical values
for simulated states A, B, C and D are compared.
\label{fig:vitpscomp}}
\end{figure*}
Fig.~\ref{fig:vitpscomp} displays the sound velocities in both types of samples, along with the 
results of Domenico~\cite{DO77}, measured on a glass bead sample in a higher pressure range. 
Within the range of vertical pressure $P$ investigated in the experiments of~\cite{JM01}, 
the packing fraction of dry beads varies between 0.633 and 0.637, while lubricated ones
are less dense, $\Phi$ ranging between 0.613 and 0.617. Sound velocities are nevertheless larger in lubricated packings.

Comparisons with our numerical data, also shown on Fig.~\ref{fig:vitpscomp},
in spite of the differences in preparation procedures (which are idealized, and
involve somewhat arbitrary choices of parameters in simulations) and loading 
(oedometric compression in experiments, isotropic compression
in simulations), reveal some interesting qualitative convergences and semi-quantitative agreements.
Specifically, we note that:
\bi
\im
Numerical ``lubricated'' samples B are also less dense, but stiffer than numerical ``vibrated'' samples C.
\im
Sound velocities in B samples increase with $P$ slower than in C ones, and sound velocities in 
lubricated laboratory samples increase slower than in vibrated ones.
\im
Numerical C samples are better models for dry experimental packings assembled by vibration than A ones: values of
elastic moduli are in much better agreement.
\ei
One may therefore attribute the difference in sound velocities reported in~\cite{JM01} between dry and
lubricated packings to a difference in coordination number, like in numerical states B and C. (Such an interpretation
differs from the one set forth in~\cite{JM01}, which relies on the filling of open interstices by the lubricant).

The traditional numerical route to obtain dense samples, \emph{i.e.} the use of a vanishing or low friction coefficient
as for systems A and B, fails to reproduce the elastic properties of dense samples assembled by vibration. Those appear to be better
simulated with the newly introduced numerical procedure resulting in C samples, which have a much lower coordination number for the
same density. Laboratory samples with a solid fraction approaching the RCP value might well, especially if their preparation involves
vibrations or tapping, possess as small a density of force-carrying contacts as our numerical samples of type C ($z\simeq 4.05$).

Such a conclusion, in favor of low-coordinated numerical samples as better models for experimental dense packings of
dry beads than conventional, A-type ones, appears to contradict the results of Makse \emph{et al.}~\cite{MGJS99,Makse04}. Those authors
simulated what we denoted as the A0 sample series, and reported good agreements with Domenico's experimental results~\cite{DO77} and with their own
measurements. We checked that the agreement between their numerical results and our A0 data was excellent. We attribute
the conflicting conclusions to their comparison being done in a much higher pressure range than the one of Jia and Mills' experiments:
as apparent on Fig.~\ref{fig:vitpscomp}, the confining pressures in Domenico's experiments are all above  2~MPa. Likewise, 
$P$ values all exceed several MPa in the experiments performed by Makse \emph{et al.}. In this range, differences between
A and C samples, as apparent on Fig.~\ref{fig:vitpscomp} for sound velocities, as well as on~\cite[Fig. 2a]{iviso2}
for coordination numbers, tend to dwindle as $P$ increases. The discrepancy between numerical results on A-type systems and experimental
results on dry bead packings is much lower under high pressure. Yet, numerical samples of type C still fit the experimental
data better. The apparent exponent in a power-law increase of sound velocities
with $P$, \emph{i.e.} the slope on Fig.~\ref{fig:vitpscomp}, is, in particular, better reproduced by C data than by A ones.
On discussing such a high pressure range, one should nevertheless keep in mind the possible occurrence of non-elastic behavior
in the contacts, as pointed out in~\cite{iviso2} (paper II), where the maximum stress levels in contact regions were estimated.

The fast increase of $G$ as a function of $P$ in C samples is not observed in the experimental results of Jia and Mills. 
Furthermore, using formula~\eqref{eqn:defnu} (which assumes isotropy), these results correspond to Poisson ratios (between 0.32 and 0.34), which
are larger than numerical results and do not vary with $P$. However, these data are bound to be affected by stress anisotropy.

\begin{figure}[!hb]
\centering
\includegraphics[angle=270,width=8.5cm]{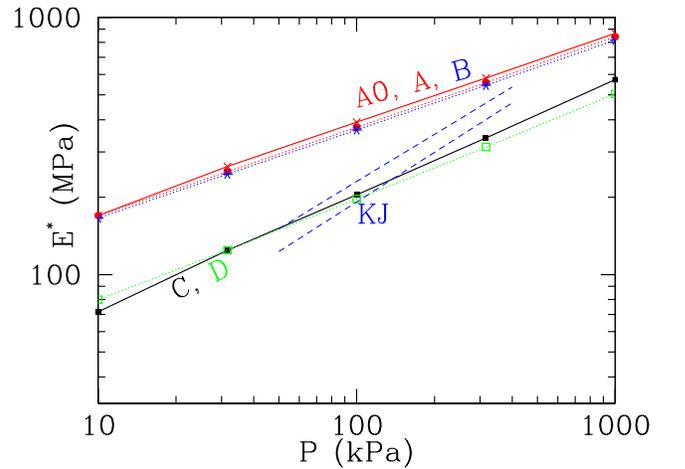}
\caption{\label{fig:mode} (Color online) 
Young modulus $E^*$ in numerical samples A, A0, B, C, D, labelled with same colors and symbols as on Fig.~\ref{fig:vitpscomp}, 
as a function of $P$, compared to fits through
data points of Kuwano and Jardine (dotted lines, KJ)~\cite{KJ02}, with two sets of values because of fabric anisotropy.}
\end{figure}

%\subsection{Gravity-deposited packings under isotropic stresses}
In this respect, 
a comparison of numerical results with the data of Kuwano and Jardine~\cite{KJ02} is easier,
 as those were measured in glass bead samples under \emph{isotropic} stress states, 
from about 50 to 400~kPa. The samples of Kuwano and Jardine have similar densities to D ones ($\Phi \simeq 0.59$), 
and are initially made by ``air pluviation'',
\emph{i.e.,} deposited under gravity, in a controlled procedure ensuring homogeneity. 
The values of shear moduli are 
close to the numerical values for C and D states, and vary with $P$ even faster, as $G\propto P^{0.55}$. The power law fit through
these data correspond to the line marked ``KJ'' on Fig.~\ref{fig:modgp}. Kuwano and Jardine, combining static small-strain tests and sound velocity 
measurements, could evaluate the 5 independent elastic moduli of the transverse isotropic granular material assembled under gravity. 
To compare our numerical results, obtained in isotropically assembled systems, with theirs, 
we ignored the moderate effect of the fabric anisotropy on experimental elastic moduli and used, in Fig.~\ref{fig:modgp}, the moduli
corresponding to a shear strain in the vertical plane, the shear modulus in the horizontal plane being about 7\% larger. 
Another similarity between our results in
D or C states and the data of \cite{KJ02} is the pressure dependence of the two Poisson ratios $\nu_{vh}$ and $\nu_{hh}$ 
which couple stress and strain components in 2 different 
directions respectively  defining vertical and horizontal planes. 
Despite some scatter in the measured values of these ratios, $\nu_{vh}$ and $\nu_{hh}$ show a marked decreasing trend
between P=80~kPa and P=400~kPa. Finally, we compare the Young moduli mesured in~\cite{KJ02} in vertical 
and horizontal directions to our numerical values in states A to D on Fig.~\ref{fig:mode}.
Our numerical values for Young modulus $E^*$ in systems with low coordination numbers are similar to those results, 
but the pressure variation seems faster (with exponent $\sim 0.6$) in experiments. 

We conclude therefore 
that, although more systematic confrontations with experimental results are necessary, some features of the
moduli in low coordination numerical packings are apparently observed in the rather loose glass bead samples of Kuwano and Jardine. 
\section{Elastic r\'egime\label{sec:elastreg}}
The elastic moduli express the relationship between small
stress and strain increments. We now wish to evaluate the elastic range, and to explore the origins of the
breakdown of elasticity for larger increments. 
\subsection{Limit of linear r\'egime}
Motivated by comparisons with experiments, we tested the
predictions of linear elasticity, as evaluated with the moduli obtained from the stiffness matrix, versus the full MD simulation for
small and slowly applied load increments, in the case of a \emph{triaxial axisymmetric compression}. 
This test, familiar
in geomechanical engineering~\cite{DMWood,MIT93,BiHi93}, consists in increasing one stress component, say $\sigma_1$, 
while the other 2 are kept constant at the
initial value of the isotropic pressure in the initial state: $\sigma_2=\sigma_3=P$. More often, in the laboratory,
one controls strain component $\epsilon_1$ (called ``axial strain'') with the motion of a piston, imposing a constant, slow strain rate
$\dot \epsilon_1$, while the lateral stresses are maintained by the pressure
of a fluid surrounding the sample, which is wrapped in an impervious membrane. It is customary to express the results of such a test with two curves, 
representing, as functions of axial strain, the \emph{deviator stress} $q=\sigma_1-\sigma_3$ and the \emph{volumetric strain}
$\epsilon_v = \epsilon_1+\epsilon_2+\epsilon_3$, the initial isotropic state being chosen as the origin of strains. 

One should have in the quasistatic r\'egime (small enough $\dot \epsilon_1$), within the linear elastic range, for small enough $\ww{\epsilon}$, 
\be
\ba
q&=E^*\epsilon_1\\
\epsilon_v &= (1-2\nu^*) \epsilon_1,
\ea  
\label{eqn:triax}
\ee
where $E^*={\disty \frac{9BG}{3B+G}}$ and $\nu^*$ (see Eqn.~\ref{eqn:defnu}) 
are respectively the Young modulus and the Poisson ratio of the material in the initial state.

\begin{figure}[!htp]
\centering
\subfigure[Deviator stress normalized by $P=\sigma_3$ versus axial strain $\epsilon_1$, 
for the five $P$ values indicated.]{
\includegraphics*[angle=270,width=8.5cm]{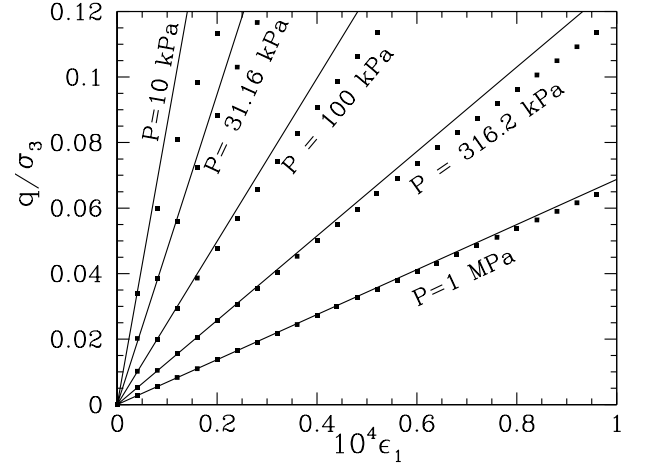}
\label{fig:prq1}}
\subfigure[Volumetric strain $-\epsilon_v$, showing contractance, versus axial strain, same
$P$ values, growing according to arrow. ]{
\includegraphics*[angle=270,width=8.5cm]{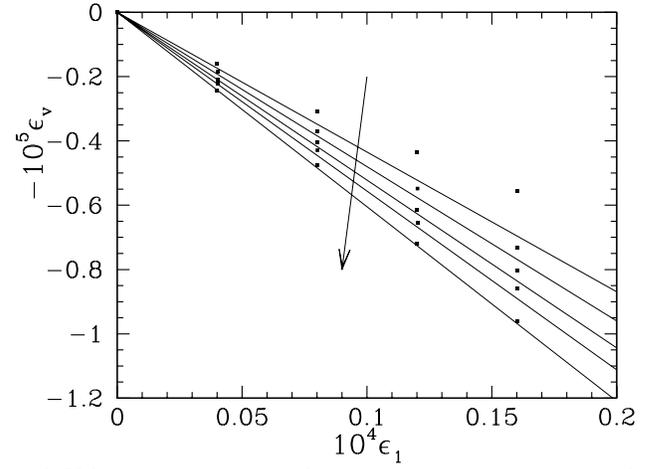}
\label{fig:prq2}}
\caption{Deviator stress (a) and volumetric strain (b)
versus axial strain in beginning of triaxial compression of a type C sample. Dots show MD results of triaxial compression while straight
lines have slopes given by~\eqref{eqn:triax}. The volumetric strain is shown with an axis oriented downwards
to better visualize the \emph{decrease} in sample volume, which contrasts with its dilatancy for larger strains.
\label{fig:limest1C}
}
\end{figure}

We simulated triaxial compressions for small axial strains and compared the resulting deviator stress and volumetric strain curves to~\eqref{eqn:triax}.
To minimize dynamical effects in simulations of quasistatic behavior, the inertia parameter $I$
 was kept below $10^{-4}$ or even  $10^{-5}$
for the most fragile, low pressure samples. 
($I$ is defined by $I = {\disty \dot \epsilon \sqrt{\frac{m}{aP}}}$ with $m$ denoting the mass of one grain, see papers I and II~\cite{iviso1,iviso2}).
Fig.~\ref{fig:limest1C} shows the typical results of such a comparison in the case of one C sample,
under isotropic initial pressures P growing from 10~kPa to 1~MPa.
This comparison first shows that full MD computation results do admit, in excellent appproximation, as slopes of the tangents to the 
curves at the origin, the appropriate values deduced from the evalution of elastic moduli by the stiffness matrix approach, 
as expressed by~\eqref{eqn:triax}. This confirms the statements made in Sec.~\ref{sec:defelastic}, and checked in Appendix~\ref{sec:appendixfnft}
about the definition of elastic moduli: this definition makes sense as a very good approximation in spite of the slight directional dependence of
contact stiffness matrix $\kk$.

Values of $\epsilon_1$ for which the $q(\epsilon_1)$ curve significantly deviates from its tangent at the origin increases from about 
$\sim 10^{-5}$ to  $\sim 10^{-4}$ as P grows from 10~kPa to 1~MPa. This linear elastic range limit, if expressed with
stress ratio $q/\sigma_3$, shows but a slow increase with $\sigma_3=P$, starting around $q/\sigma_3\simeq 0.05$.
The initial slope of the volumetric strain curve 
increases with pressure, in C samples, in agreement with the results on Poisson ratios (Sec.~\ref{sec:elastpnum}). Deviations from the linear range
of~\eqref{eqn:triax} appear a little sooner, for $\epsilon_1$ slightly below $10^{-6}$ at 10~kPa, increasing to a few times $10^{-5}$ at 1~MPa.

\begin{figure}[!htp]
\centering
\includegraphics*[angle=270,width=8cm]{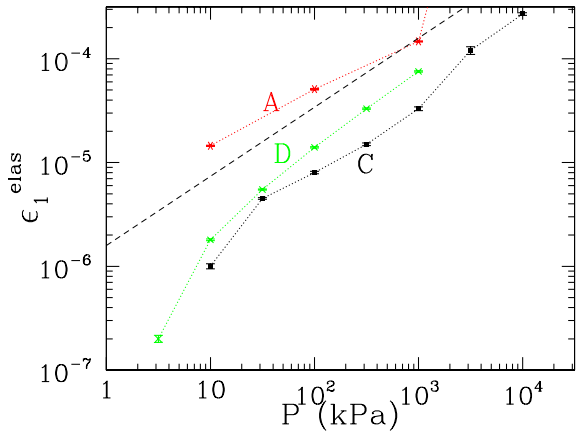}
\caption{\label{fig:limest1} (Color online) Threshold $\epsilon _1^{\text{elas}}$ above which $q$ differs from $E^*\epsilon _1$ by more than $5\%$, versus $P$, for
series A (red), C (black) and D (green). The dashed line has slope $2/3$. Data points corresponding to one sample series are connected by thin dotted lines.}
\end{figure}

In the literature on sand properties~\cite{Tat104,HI96,KJ02}, 
it is often observed that the approximately linear elastic range about a prestressed reference state extends to
strains of order $10^{-6}$ or $10^{-5}$. On Fig.~\ref{fig:limest1}, we plotted the value of $\epsilon_1$ for which the deviator stress starts to deviate
from \eqref{eqn:triax} by more than $5\%$, versus the confining pressure, for series A, C, and D.
Recalling that most experimental results are collected in the range $50$~kPa$\le P\le 1$~MPa, 
these data confirm the experimental observations~\cite{Tat104} on sands 
in terms of order of magnitudes for all three sample series. However, they also witness a systematic 
growth of the elastic threshold $\epsilon_1^{\text{elas}}$ with $P$, roughly as $P^{2/3}$. This suggests a constant elastic deviator interval 
relative to the confining pressure, $q^{\text{elas}}/P$, on assuming $E\propto P^{1/3}$.  Figs.~\ref{fig:limest2A} and \ref{fig:limest2CD} show that
the elastic range is better expressed in that form, as the threshold ratio $q^{\text{elas}}/P$ displays much smaller variations as a function of $P$:
\begin{figure}[!htp]
\centering
\includegraphics*[angle=270,width=8cm]{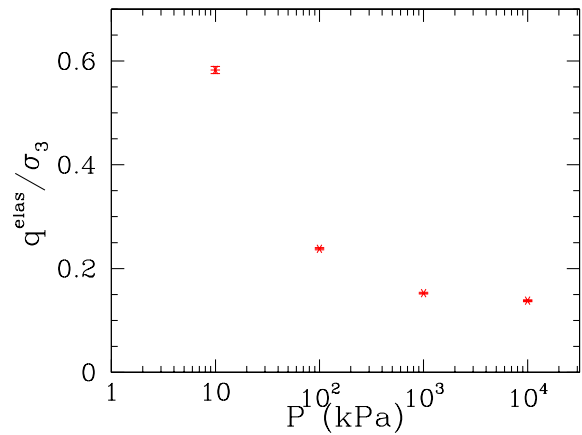}
\caption{\label{fig:limest2A} Stress ratio $q^{\text{elas}}/P$ above which $q/E^*$ differs from $\epsilon _1$ by more than $5\%$, versus $P$, for
series A.}
\end{figure}
unlike Fig.~\ref{fig:limest1}, those graphs show stress intervals on a linear scale. Expressing
the extension of the linear elastic r\'egime in terms of strains, as done in the literature on sands,
allows one to gather the different sample series within the same range of magnitudes around $10^{-5}$, provided
the confining pressure stays within the interval that is most often investigated in experiments. However, the pressure dependence is better accounted for on
expressing the upper limit of the linear elastic range in terms of stress increments, relative to the confining stresses.
The trend in low-coordinated systems C and D is an
\begin{figure}[!htp]
\centering
\includegraphics*[angle=270,width=8cm]{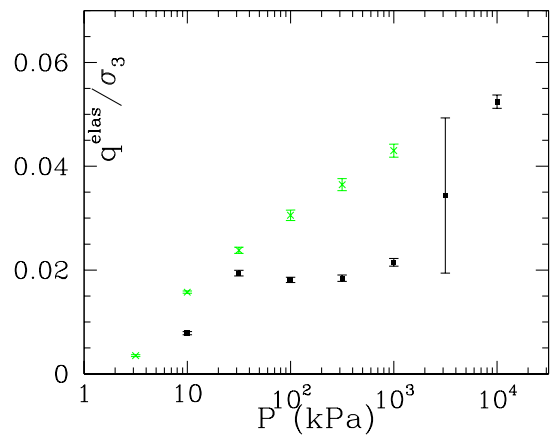}
\caption{\label{fig:limest2CD} (Color online)
Stress ratio $q^{\text{elas}}/P$ above which $q/E^*$ differs from $\epsilon _1$ by more than $5\%$, versus $P$, for
series C (black, lower data points)  and D (green).}
\end{figure}
increase of the linear elastic interval, expressed as a stress ratio, with $P$. At the lowest pressure, the smallness of this interval (typically about $10^{-3}$ 
for ratio $q/P$ in D samples under $P=1$~kPa) is characteristic of the greater fragility of tenuous networks, a phenomenon, once again, reminiscent of the situation of
nearly isostatic force-carrying structures in frictionless packings under low pressure. 
In the limit of rigid grains, frictionless systems (if their properties
do not qualitatively depend on space dimension)
should behave as described in~\cite{CR2000}: in packings of rigid disks the deviator
stress increments causing exactly isostatic 2D contact networks to fail and rearrange were found to approach zero as an inverse 
power of the number of particles in the limit of large systems.

Series A configurations, on the other hand, have decreasing linear elastic intervals as a function of pressure. This is due to
the increase of friction mobilization, which leads to larger non-elastic terms in the response to load increments. 
\subsection{Onset of irreversibility}
\begin{figure}
\includegraphics[width=8.5cm]{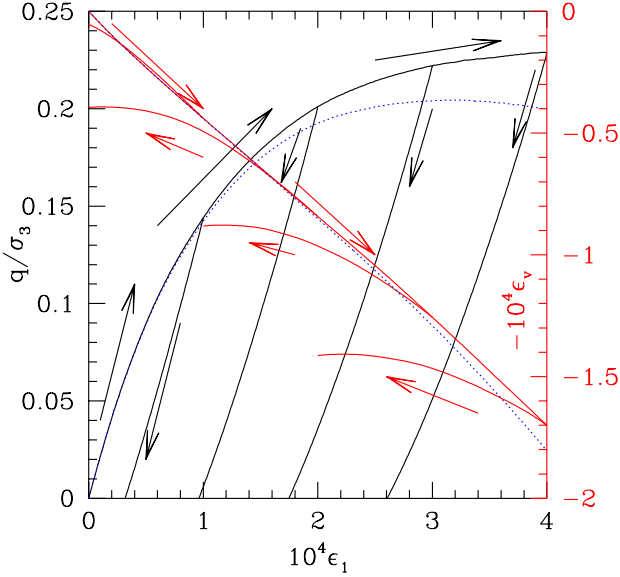}
\caption{\label{fig:triaxrev}
(Color online) Deviator stress (black, left axis) and volumetric strain (red, right axis)
versus axial strain in triaxial compression for small strains, with unloading curves, in a D sample initially under P=100~kPa. The blue dotted curves show the results of 
a calculation with the sole contacts that are initially present.}
\end{figure}

\begin{figure}
\includegraphics*[angle=270,width=8.5cm]{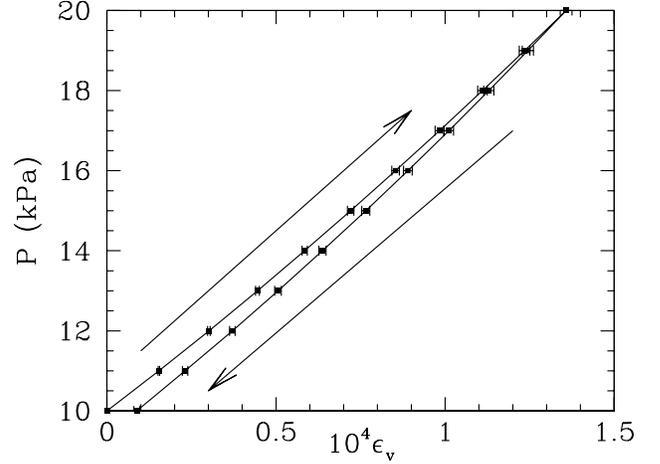}
\caption{\label{fig:compiso2}
Isotropic pressure increase versus relative volume change in pressure cycle, $P$ growing from 10 to 20~kPa in C samples, and then decreasing back to 10~kPa.}
\end{figure}

So far we have been testing the accuracy of \emph{linear} elasticity, \emph{i.e.,} the predictions of~\eqref{eqn:triax} with the initial moduli. 
It is known from experiments that granular materials cease to be elastic outside this linear r\'egime. This may be checked
on testing for reversibility in a strain cycle. 
The effects of unloading from various points on the triaxial
compression curve are shown on Fig.~\ref{fig:triaxrev}, in a type D sample with $\sigma_2=\sigma_3=100$~kPa. 
On reversing the sign of $\dot \epsilon _1$, while still maintaining 
constant lateral stresses $\sigma_2=\sigma_3$, one observes that the unloading curve starts with a slope close to the initial slope of the loading curve. This
common slope is the elastic moduli corresponding to the linear response for very small strain increments. Consequently,
the response to a deviator stress is no longer reversible as soon as it ceases to be linear, as shown on Fig.~\ref{fig:triaxrev}, on which the axial strain that results from 
a deviator stress cycle is represented. A large
proportion of the strain entailed by a relative stress increase of order $10^{-1}$ is not recovered on returning to the initial load. 
The response to an isotropic load increment is much closer to reversibility,
 even for much larger stress variations, as shown on Fig.~\ref{fig:compiso2}, which displays 
the stress-strain curve in an isotropic pressure cycle. As $P$ varies by a factor of 2, about 
93\% of the volume increase is recoverable upon unloading to the initial pressure. 
Only for the large pressure cycles as studied in paper II~\cite[Fig. 2]{iviso2}
can one observe notable irreversible changes, in coordination number rather than in density. One thus finds again
that the response to increments in load \emph{intensity} (here: isotropic compression) strongly differs 
from the response to changes in load \emph{direction} (here: deviator stress).
\subsection{Origins of nonlinearity and irreversibility}
\begin{figure}
\includegraphics[width=8.5cm]{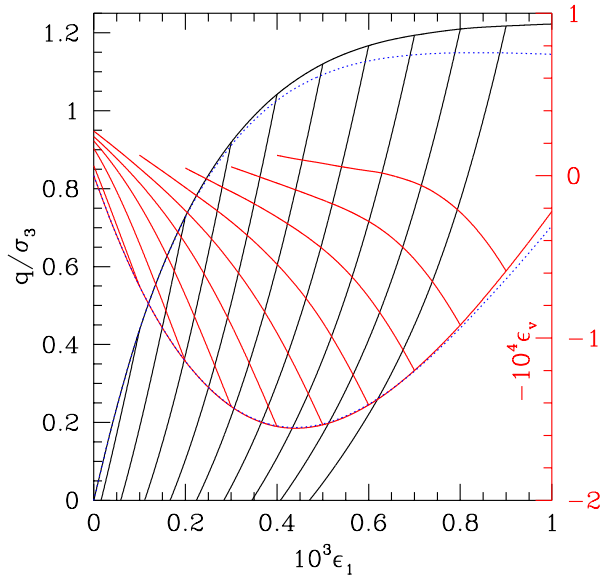}
\caption{\label{fig:triaxrevA}
(Color online) Analog of Fig.~\ref{fig:triaxrev} in a sample of type A. Note the different scales.}
\end{figure}

For deviatoric stress increments, the gradual onset of irreversibility, as the applied load variation increases,
coincides with the breakdown of linearity. The lack of reversibility, as shown 
on Fig.~\ref{fig:triaxrev}, has two different origins. One is the mobilization of friction, and the second is the failure of the contact network: 
the packing eventually breaks apart and rearranges. 
In order to detect the occurrence of this latter kind of event, one may compute the response
in the beginning of a triaxial compression of some samples with MD calculations in which only the initially present contacts,
in the isotropic state, are
taken into account. One thus tests the ability of the initial contact network to support different stress values. One then
observes, as shown on Fig.~\ref{fig:triaxrev}, that the initial contact network proves unable to support a deviator stress
beyond a certain limit: the $q$ versus $\epsilon_1$ curve reaches a maximum if $\dot\epsilon_1$ is controlled. 
This witnesses the propensity of
the packing to become unstable and gain kinetic energy before it finds a new contact network that is able to support a larger
deviator. This happens the
sooner the larger the stiffness parameter $\kappa$ in systems with low coordination numbers. Thus the corresponding 
ratio $q/\sigma_3$ decreases from about 0.2 in the example of Fig.~\ref{fig:triaxrev}, corresponding to one D-type sample under 100~kPa,
to, typically, 0.1 under 3~kPa, and 0.05 under 1~kPa. In this respect (as for the values and pressure dependence of shear moduli) 
low coordination frictional packings exhibit, 
in a weakened form, similar singular behaviors to frictionless ones. 
The opposite behavior is that of A-type packings, with a very large coordination number. As shown on Fig.~\ref{fig:triaxrevA}, 
the irreversibilities are smaller and the initial contact network proves able to withstand considerably larger relative deviator
stresses,  $q/\sigma_3\sim 1.1$ under 100~kPa.
The behavior of A packings under shear is therefore somewhat intermediate 
between that of low coordination systems C or D under not too low confining stress, and the response to 
isotropic pressure increases. The behavior of A samples under triaxial compression is to a large extent determined 
by the response of the initial contact network, and the rise of
deviator $q$ as a function of axial strain is very fast. This is a typical feature of well-coordinated packings, 
as studied in~\cite{Roux05}. In Refs.~\cite{Roux05} and \cite{RC02}, 
two different types of strain are distinguished: those due to the deformability of contacts, 
and those stemming from network failures and rearrangements. As long as the
first type of strains is the only one present, the behavior is close to that of the inital contact network. 
Beyond the stability range of the initial contact network, 
the effect of rearrangements dominate~\cite{RC02}, and
strains are produced by local instabilities which can be described with a rigid grain model~\cite{Staron02b}.

Refs.~\cite{Gael2,RC02} show that well-coordinated isotropic packings (2D analogs of 
A systems) can support rather large deviator stress increments in the $\kappa\to \infty$ limit, whatever the sample size. 
Systems with lower coordination numbers appear to exhibit
intermediate behaviors between this one and the ``fragility'', defined as the propensity to rearrange for
arbitrary small stress increments in the large system limit~\cite{CWBC98,JNR2000,CR2000,RC02}, 
of assemblies of rigid, frictionless grains.  The stability range of given equilibrium contact networks 
extends to smaller stress increment intervals in C or D-like packings, but we expect it to remain finite for arbitrary low pressure levels. 
These properties, and the distinction of two types
of strains, are further explored and discussed in a forthcoming paper~\cite{CR07}.
\section{Conclusions and perspectives\label{sec:conc}}
Our numerical results can be summarized as follows.

Elastic moduli of granular packings are primarily sensitive to the stress level, \emph{via} the average contact stiffness, which is proportional
to $P^{1/3} (z\Phi)^{-1/3}$ under pressure $P$. Once this effect is taken into account, important differences remain between the elastic properties
of different packing structures, and systems assembled with the same density might exhibit large variations in their moduli, since those are essentially related to
coordination numbers. Under isotropic pressures, one should distinguish between bulk and shear moduli. 
The bulk modulus, in all studied configurations, is efficiently
evaluated by the Voigt and Reuss-like bounds, with an error smaller than 20\%. 
In general, we expect a difference between the responses to changes in 
load intensity on the one hand, and to changes in stress direction on the other hand.
In isotropic systems, the latter correspond to purely deviatoric stresses, the effect of which differs the most from that of hydrostatic pressure increments in
low coordination contact networks. In such cases shear moduli are anomalously small and increase faster with the confining pressure.  
In well coordinated states, such as A and B, satisfactory estimates of $B$ and $G$ moduli on using 
improvements of the Voigt approximation, based on locally independent fluctuations about average strains, such as the La Ragione-Jenkins (LRJ) approach. Moreover,
the additional stiffening effect of the increase in coordination number, due to compression, might, 
for pressures in the MPa range, be reasonably predicted with the homogeneous shrinking assumption, or similar refinements thereof.
Such schemes nevertheless require rather detailed statistical knowledge of local particle configurations. On the other hand, the shear response of
low coordination packings, such as C and D, is 
better described with reference to a state with no force indeterminacy, even though such a state is not closely approached in the limit of low pressures (or large
stiffness level, $\kappa\to +\infty$)
except for unphysically large friction coefficients (as in the Z configuration series).  
In the rigid, $\kappa\to +\infty$ limit, shear moduli become proportional to the level of force indeterminacy, which
directly relates to $z^*-4$, with a small correction due to divalent particles (hence the definition and use of $z^{**}$). The dependence of $z^*-4$ 
on pressure seems however difficult to predict with the necessary accuracy~\cite{iviso2}.
The physical origin of the breakdown of linear elasticity beyond an interval of very small strains depends on $z^*$ as well. 
Linear elasticity fails considerably
later, in terms of relative stress increments, for isotropic pressure changes, and, to a lesser extent, 
in high coordination packings subjected to deviatoric stress paths. In such cases linear elasticity breaks down because of frictional forces. 
Elasticity ceases to apply for very small shear stress increments in low coordination systems, the smaller the closer they are to 
packings devoid of force indeterminacy. 
In such cases the main physical origin of nonelastic response is network fragility, as the contact structure breaks apart in response to stress increments. 
Extension of linear elastic r\'egimes observed in numerical simulations 
agree semi-quantitatively with observations on sands. 
The shape of the stress-strain curves beyond the elastic range correlates with the coordination number,
with a much stiffer response in well coordinated packings. 
On comparing numerical and experimental results, the low pressure r\'egime of poorly coordinated networks corresponds to the lowest pressures for which 
laboratory results are available.
The characteristic features of this r\'egime, such as $G$ increasing with $P$ faster than $B$, or $V_S$ faster than $V_P$, 
as apparent in C and D numerical configuration series,
are not observed in the experimental data of Jia and Mills~\cite{JM01,ARMJM05} on dense
samples. Yet some other measurements made by 
Kuwano and Jardine on looser sphere packings~\cite{KJ02} show similar trends to C and D-state simulations.
 
The variety of observations corresponding to the same pressure and density values for the same material confirms 
the sensitivity of elastic moduli to otherwise undetectable differences in
inner structures.  It seems in particular, although information about the full stress tensor in the measurements of~\cite{JM01,ARMJM05} is lacking, that
packings densified by vibrations or repeated shakes have a smaller coordination number than lubricated ones, 
in spite of a possibly larger density. Additional experimental results with more detailed information on stresses and anisotropy of elastic moduli in packings
assembled by different procedures, as well as simulations of anisotropic packings, could enable more quantitative comparisons.

Based on those results, several interesting perspectives should be pursued in the near future.

On the theoretical side, it seems promising to study how granular packings, within and outside the elastic range, 
deform and destabilize, in more microscopic detail. 
Basically, packings with few contacts are closer to failure, and some of their anomalies in elastic properties correlate with failure mechanisms.
Amorphous systems made of model atoms or molecules at zero temperature have been characterized, in this respect, 
in terms of an intrinsic scale~\cite{TWLB02,LBTWB05}, and elementary plastic rearrangement events, the spatial structure of which is similar to
that of nonaffine elastic displacements, have been investigated~\cite{MaLe06}. Granular materials have friction, which requires
more sophisticated stability analyses of given contact networks~\cite{RC02,KuCh06,Bagi07,McHe06}, 
interact with much stiffer force laws, and exhibit characteristic dilatancy properties and fabric evolutions under strain. 
It is worth investigating in greater detail the
possible similarities and differences between their quasistatic rheology and the plasticity of amorphous materials.
A characteristic length scale, which diverges in the isostatic limit,
has also been invoked in relation with the singular elastic properties of frictionless packings~\cite{WNW05}.
It should be examined whether such a length plays a role in nonelastic deformation behaviors, 
and similar investigations should be carried out in systems with friction, 
which also exhibit complex, long-range correlated strain fields~\cite{Radjai02}. 

More practical issues which deserve investigations are how elastic moduli, which can be measured in equilibrated packings under varying stresses~\cite{GBDC03}, 
can be used to infer useful information
on their inner structure, which, in turn, can be exploited to predict their behavior under larger disturbances. 
As an example, the coordination number of an isotropic packing, if it is large,
will result in a stiff response to deviator stress increments, characteristic of a stable contact network, and
a faster mobilization of internal (macroscopic) friction as a function of strain (see Fig.~\ref{fig:triaxrevA} and Ref.~\cite{Roux05}). 
Numerical simulations of
anisotropic stress and fabric states, of stress paths and large strains, and further numerically based correlations 
between elastic properties and stresses, strains and inner structures
are of course necessary. Finally, the geometry of polydisperse and non-spherical particles should also be explored. 
Such packings might have large populations of rattlers, some
more collective stable floppy modes than in the case of spheres~\cite{Donev-ellipse}, and different mechanical properties~\cite{Matsu05}.

\appendix
\section{Tangential elasticity and friction \label{sec:appendixfnft}}
We investigate here the effect on elastic moduli of the corrections to the contact law advocated in~\cite{EB96} that we
use in our simulations, and we also discuss the possible effects of more sophisticated models, in which
the partial mobilization of friction and the presence of a sliding region within the contact area~\cite{MiDe53,JO85} is taken into account.

Strictly speaking, those terms preclude the definition of a perfectly elastic response, which should be reversible and involve a uniquely
defined stiffness matrix.
 
The simplified law we adopt
involves a tangential stiffness $K_T$ depending on the normal deflection $h$, but independent of the current
mobilization of friction. This is the same
approximation as used in~\cite{MGJS99,Makse04}: the value of $K_T$ is the correct one in the absence of
elastic relative tangential displacement, when ${\bf T}=0$. 

The rescaling of $K_T$ we chose to apply in situations of decreasing normal force $N$, in order to avoid energetic inconsistencies, as
explained in paper I~\cite{iviso1}, means that the contact stiffness matrix, as defined in Section~\ref{sec:rist}, should then be 
written differently. Its block corresponding
to contact $i,j$ is not given by $\kk _{ij}^E$, as written in~\eqref{eqn:kiju}, but takes another form $\kk _{ij}^R$ for a receding pair of grains.
Since $K_T/K_N$ is constant, and $K_N\propto N^{1/3}$, one has, with ${\bf n}_{ij}$ and ${\bf T}_{ij}/\normm{{\bf T}_{ij}}$ as
first and second basis vectors,
\be
\kk _{ij}^R = \bma K_N(h_{ij}) & 0 & 0 \\ 
\frac {\strut \normm{\Tij }}{\strut 3N_{ij}}K_N(h_{ij}) &K_T(h_{ij})& 0 \\ 0&0&K_T(h_{ij})
\ema
.
\label{eqn:kijr}
\ee
The non-diagonal element of~\eqref{eqn:kijr} is smaller than $\mu K_N(h)/3$ ($K_N/10$ for $\mu=0.3$), 
and its effects are likely limited if friction is not too strongly mobilized, as should be the case under isotropic loads. 
Let us denote as $\Delta \sti$ 
the correction to the symmetric form of stiffness matrix 
$\stia \simeq \sti $ (see Section~\ref{sec:rist}), in which Eqn.~\eqref{eqn:kiju}
is applied to all contacts, due to this treatment of decreasing normal forces. 
We solved the linear system of equations~\eqref{eqn:defsti}, with loads corresponding to different global stress increments, 
to first order in the perturbation $\Delta \sti$:
$$
{\bf U} \simeq {\bf U}^{(0)} + \Delta {\bf U},
$$
where ${\bf U}^{(0)}$ is the solution to the unperturbed problem, ${\bf U}^{(0)} = \sti ^{-1} \cdot \fext$,
and $\Delta {\bf U}$ is the correction:
\be
\Delta {\bf U}=- \sti ^{-1}\cdot \Delta \sti \cdot {\bf U}^{(0)}.
\label{eqn:sol1}
\ee
In~\eqref{eqn:sol1} one should pay attention to 
the directional dependence of $\Delta \sti$  (according to whether $h$ increases or decreases). 
The first-order correction is therefore not linear, but depends linearly on the amplitude of the load increment 
with a coefficient depending on its direction.
We evaluated the resulting correction to the compliance in the cases of
uniaxial (\emph{e.g.}, $\Delta \sigma _1 >0$ or  $\Delta \sigma _1 <0$ while
$\Delta \sigma _2 = \Delta \sigma _3=0$), isotropic (positive or negative value of $\Delta \sigma _1 =
\Delta \sigma _2 = \Delta \sigma _3$) and purely deviatoric (\emph{e.g.}, $\Delta \sigma_1 =-\Delta \sigma _2 $ and
$\Delta \sigma _3=0$) stress increments. Relative corrections never exceeded $1\%$, 
the largest ones, as expected, being observed for an
isotropic pressure \emph{reduction} (which tends to reduce normal contact forces).  

Our contact model also introduces an approximation, which we now discuss. As opposed to 
more sophisticated implementations of the contact law, as used by some authors~\cite{THRA88,TH00}, our model does not
keep track of the local slip
distribution within the contact region. The \emph{maximum} effect of such slip is to \emph{reduce}
 the tangential stiffness from $K_T(N)$ to
\be
K'_T(N,{\bf T}) = K_T(N) (1 - \frac{\vert\vert {\bf T}\vert\vert}{N})^{1/3}.
\label{eqn:ktnt}
\ee
in the ``loading'' direction (\emph{i.e.,} tending to increase $\norm{T}/N$), 
and for the tangential relative displacements along ${\bf T}$.
The possible influence on the simulated elastic properties
of our overestimating the tangential stiffness of the contacts was assessed as follows. 
We computed the elastic moduli for the equilibrated configurations, keeping the same values of 
contacts forces, using formula~\ref{eqn:ktnt} for all contacts, in both
tangential directions. (Such a calculation thus implicitly assumes that
the equilibrium force distribution is not affected by the change in the contact law). This procedure
exaggerates the effects of slip and gradual friction mobilization: formula~\ref{eqn:ktnt} gives the lowest
possible
value for $K_T$ and only applies in specific loading histories, and for stress increments that tend to increase
$\frac{\norm{T}}{N}$. We found that the relative corrections to computed elastic moduli evaluated with this procedure never exceed 3\%.

Consequently, it is a very good approximation to replace the contact stiffness matrix $\kk$ by its diagonal form given
by Eqn.~\eqref{eqn:kiju}, provided friction is not fully mobilized in any contact.

If condition $\norm{T}=\mu N$ is reached in contact $i,j$, then matrix $\kk _{ij}$ has to be written as follows.
With the same choice of basis vectors as for~\eqref{eqn:kijr}, $\kk _{ij}$ has a ``loading'' form $\kk _{ij}^L$  given by
\be
\kk _{ij}^L = \bma K_N(h_{ij}) & 0 & 0 \\ \mu  K_N(h_{ij}) & 0 & 0 \\ 0&0&K_T(h_{ij})
\ema
\label{eqn:kijl}
\ee
and an ``unloading'' one equal to $\kk _{ij}^E$ or to $\kk _{ij}^R$, 
depending on whether $\delta u_N$ is increasing or decreasing.
If it is increasing, the loading form $\kk _{ij}^L$ should be used 
if 
$$K_T(h_{ij})\delta {\bf u}^T_{ij} \cdot \frac{{\bf T}_{ij}}{\normm{{\bf T}_{ij}}} -
\mu K_N(h_{ij}) \delta{\bf u}^N_{ij} >0.$$
If $\delta u_N$ is decreasing, this condition becomes
$$K_T(h_{ij})\delta {\bf u}^T_{ij}\cdot \frac{{\bf T}_{ij}}{\normm{{\bf T}_{ij}}}
+\left[\frac{\normm{{\bf T}_{ij}}}{3N_{ij}}-\mu \right] K_N(h_{ij}) \delta{\bf u}^N_{ij}>0.
$$
Note that $\kk _{ij}^L$ is a non-symmetric singular matrix of rank 2. As remarked in Section~\ref{sec:defelastic}, well-equilibrated
configurations prepared by molecular dynamics do not contain any contact where the condition $\norm{T}=\mu N$ is exactly reached.
This justifies our using the diagonal form $\kk _{ij}^E$ of the local stiffness matrix block pertaining to any contact.
Full mobilization of friction occurs under small load increments, and the resulting deviation from elastic behavior 
is investigated in Section~\ref{sec:elastreg}.                             

\section{Geometric stiffness matrix and treatment of divalent beads\label{sec:appendixTrot}}
The geometric term added to the change in intergranular forces entailed by 
small displacements and rotations was evaluated in
paper I~\cite{iviso1}. Here, for completeness, we write down its contribution to the geometric stiffness matrix $\stib$.
Those terms stem from the change in the direction of previous contact forces. They were carefully evaluated in general situations of particles
of arbitrary shapes by Kuhn and Chang~\cite{KuCh06}, and by Bagi~\cite{Bagi07}. This results in formulae
of considerable complexity, involving ``branch vectors'', joining contact points to particle centers where moments are evaluated, 
as well as particle surface curvatures,
which determine the small changes in normal directions caused by particle displacements and rotations. Those results assume much simpler 
forms in the case of spherical grains, for which both branch vectors
and radii of curvatures are given by particle radii, and our results agree with such particular forms of the general expressions of Refs.~\cite{KuCh06,Bagi07}.

In the matrix block Our results agree with the general expressions written down in those references, but  radii of sphere $i$ is simply denoted as $R_i$ below 
(in our numerical computations, all $R_i$ values are equal to $a/2$). 

The $6\times 6$ block $\stib _{ii}$ of the geometric stiffness matrix is a sum over the contacts of grain $i$, 
$$
\stib _{ii} = \sum_{j\ne i}\ww{L}_{ij},
$$
each term being given by the expressions written in paper I~\cite[appendix B]{iviso1}. Using 
a system of coordinates with $\nij$ and $\Tij$ setting the orientations of the two first axes, one has:
\be
\ww{L}_{ij}=
\bma
0& {\disty\frac{T_{ij}}{r_{ij}}} &0 &0 &0 &\phantom{{\disty\frac{A}{2}}}0\\
0&-{\disty\frac{N_{ij}}{r_{ij}}}&0&0&0&\phantom{{\disty\frac{A}{2}}}0\\
0&0&-{\disty\frac{N_{ij}}{r_{ij}}}&{\disty\frac{T_{ij}}{2}}&0&\phantom{{\disty\frac{A}{2}}}0\\
0&0&0&0&0&\phantom{{\disty\frac{A}{2}}}0\\
0&0&{\disty\frac{N_{ij}R_i}{r_{ij}}}&-{\disty\frac{R_iT_{ij}}{2}}&0&\phantom{{\disty\frac{A}{2}}}0\\
0&-{\disty\frac{N_{ij}R_i}{r_{ij}}}&0&0&0&\phantom{{\disty\frac{A}{2}}}0\\
\ema
\ee
As to the non-diagonal block $\stib _{ij}$, it is obtained from $\ww{L}_{ij}$ on reversing the signs of the coefficients in the three first columns.
Matrix $\stib$ is therefore clearly not symmetric, which, in principle, forbids the definition of an elastic energy. However, 
 each term involving $N_{ij}$ or ${\bf T}_{ij}$ in $\stib$ is negligible once compared to its counterpart in
$\stia$,  where forces are replaced by terms of order $K_N a$. 
Generally, $\stib$ terms are therefore 
of relative order $\kappa^{-1}$ if compared to the corresponding ones in $\stia$. Thus the geometric stiffness matrix only plays a role
for those directions of displacement vectors belonging to the null space of $\stia$. This is important for frictionless grains, in which 
case such floppy modes of the constitutive matrix are necessarily unstable for spheres, but not so for, \emph{e.g.}, ellipsoids~\cite{JNR2000,Donev-ellipse}.
In the case of frictional spheres we did not obtain any floppy mode on the backbone, except for beads with two contacts. However, the corresponding free motion
was shown to create no instability~\cite{iviso1}. In order to work with a positive definite stiffness matrix, as rendered necessary by numerical techniques like
Cholesky factorizations or conjugate gradient iterations, 
a stiffness term is added which associates an elastic energy with the free motion of divalent beads (the same trick is used
to impede the global translations of all grains as one rigid body). Those particles consequently have no
rotation in the direction defined by their two contact points, and the solution of the system of linear equations defined by~\eqref{eqn:defsti} is unique.

With this precaution we can therefore safely neglect the geometric stiffness matrix in all cases studied in the present numerical work.
\section{Elasticity of a prestressed system\label{sec:appendixcc}}
We  
recall some basic properties which lead to a distinction to be made, whenever
the reference configuration of an elastic continuum is prestressed, between the elastic moduli and the coefficients of the linear relation
between the Cauchy stress tensor and the strains. We specify here which type of elastic coefficients we compute in simulations.
Then, we also point out that some small 
corrections --negligible in practice -- should in principle be applied to the computed moduli, because of preexisting stresses.

Let us consider a uniform displacement gradient $\nabla {\bf u}$ within an elastic continuous medium
(derivatives with respect to coordinates on the reference, undisturbed configuration). 
In linear elasticity, the free energy density, $A/\Omega_0$, evaluated in a reference configuration (with volume $\Omega_0$), is
a quadratic function of the Green-Lagrange strain tensor $\ww{e}$, which expresses
material deformation~\cite{iviso1}. The first-order term is written with the Piola-Kirchoff stress tensor $\pk_0$~\cite{SAL01} in the reference
configuration:
\be
\Pp = \Omega_0 \pk_0 : \ww{e}.
\label{eqn:pp}
\ee 

To second order, the free energy associated with small strains involves the tensor of elastic constants $\cc$:
\be
A = \Omega _0\left[ \pk _0 : \ww{e} + \frac{1}{2}  \ww{e} : \cc :  \ww{e} \right].
\label{eqn:Ae}
\ee
Elastic moduli thus appear in the increment of $\pk$:
\be
\pk -  \pk _0 =  \cc :  \ww{e}.
\label{eqn:dpi}
\ee
The Voigt symmetry $C_{\alpha \beta \gamma \delta} = C_{\gamma \delta \alpha \beta }$
is satisfied  by the coefficients of this linear law because they are second order derivatives.
Let $\ww{L}$ denote the diagonal matrix containing the cell dimensions along the three axes, and  $\ww{L}_0$ 
its value in the reference configuration, corresponding volumes being $\Omega$ and $\Omega _0$.
With the notation $\wF = \ww{1} + \nabla {\bf u} = \ww{L}\cdot \ww{L}_0^{-1}$ 
one has $\Omega/\Omega_0=\det \ww{F} $ for
the dilation and $\ww{\sigma}$ is related to $\pk$ by 
\be
\pk = (\det \ww{F})\,  \wF ^{-1} \cdot \ww{\sigma}\cdot\trs{F} ^{-1}.
\label{eqn:relsp}
\ee
(This relation between the Piola-Kirchoff stress tensor and the Cauchy stress tensor can be found in continuum mechanics textbooks~\cite{SAL01}
and was recalled in paper I). 

 Let us now use~\eqref{eqn:dpi} and~\eqref{eqn:relsp} to write the increment of the Cauchy stress
tensor to first order in displacement gradient $\ww{\nabla  u}$ (needless to distinguish spatial derivatives in
the reference or the deformed configuration at this stage, as this would introduce second order 
corrections). Defining the linearized strain tensor $\ww{\epsilon}$ as
$$
\ww{\epsilon} = -\frac{1}{2} \left(\ww{\nabla u} + \trs{\nabla u}\right),
$$
one obtains:
\be
\ww{\sigma} = \pk_0 -\left(\tr \ww{\epsilon}\right) \pk_0 - \ww{\nabla u} \cdot \pk_0 -  \pk_0 \cdot \trs{\nabla u}
+ \cc : \ww{\epsilon}.
\label{eqn:relse}
\ee
Therefore, $\ww{\sigma}$ is not necessarily a function of the symmetric part of $\ww{\nabla u}$
only. A rigid rotation (for which $\ww{\nabla u}$ is antisymmetric) might produce a Cauchy stress increment if 
$\pk_0 $ and $\ww{\nabla u}$ do not commute. Likewise, the coefficients expressing the linear dependence of
$\ww{\sigma}$ on $\ww{\nabla u}$ do not always satisfy the Voigt symmetry, and hence one cannot regard a constant
$\ww{\sigma}$ as deriving from a potential energy of external loading. Both conditions, symmetry and dependence on
$\ww{\epsilon}$ only, are however restored if one restricts to symmetric displacement gradients, or if $\pk$
and $\ww{\epsilon}$ share common principal directions. This is always the case with our choice of boundary
conditions, or in general if $\pk _0=P\ww{1}$ is an isotropic tensor. In this latter case, \eqref{eqn:relse} relates
$\ww{\sigma}$ to $\ww{\epsilon}$, assuming isotropy of the material, with a tensor of ``apparent'' elastic
moduli $\ww{\ww{B}}$, that has the same symmetries as $\cc$. $\cc$, in isotropic systems, can be written with a bulk modulus $B$ and a 
shear modulus $G$. On relating $\ww{\sigma}$ to $\ww{\epsilon}$, the apparent moduli (as measured in an experiment)
are $B+P/3$ and $G-P$. 

It should be specified that our procedure
to compute elastic moduli in isotropic sphere packings is based on a formula for the Cauchy stress tensor. Therefore, the resulting moduli are the elements of
matrix  $\ww{\ww{B}}$, rather than $\cc$. 

As a consequence of the stress and forces that preexist
in the initial configuration before elastic response is probed, our results should also in principle be slightly
modified. As we use it, the equilibrium equation for stress components actually gives the increment of the product $\Omega \sigma _{\alpha}$,
from which the contribution $\Delta \Omega \sigma ^0_{\alpha}$, due to volume change $\Delta \Omega = -\Omega _0  \tr \ww{\epsilon}$
should be subtracted before dividing by $\Omega _0$ if the Cauchy stress variation is to be obtained. As a consequence of this correction, we should
add $P/3$ to the value of $B$ obtained with our calculation procedure. This is a very small effect, which we
have been neglecting (moduli are in MPa for stresses in kPa, see Fig.~\ref{fig:modbgp}).

\section{Voigt and Reuss bounds for elastic moduli in a sphere packing \label{sec:appvar}}
Within the approximation that the stiffness matrix does not 
depend on the direction of the stress (or strain) increment, and is symmetric (see Section~\ref{sec:defelastic} and
Appendix~\ref{sec:appendixfnft}), which fortunately proves accurate (see also Section~\ref{sec:elastreg}), the elastic r\'egime can 
be defined, and the variational properties~\eqref{eqn:min1} and~\eqref{eqn:min2} can be used. 
Variational properties leading to bounds for moduli are seldom invoked in the context of granular materials.
Our purpose here is to recall how these useful properties are established and interrelated, and how they can be exploited. 

 Let us first briefly establish the minimization property for contact force increments, 
which is less familiar than the simple minimization of elastic energy~\eqref{eqn:min1}.
We consider the solution $\Delta {\bf f}^*$ to the problem of minimizing~\eqref{eqn:min2}
among contact force increment vectors that balance the applied load increment, \emph{i.e.} such that
\be
\rigt \cdot \Delta{\bf f} = \Delta {\bf F}^{\text{ext}}
\label{eqn:consteq}
\ee
This solution is characterized by the existence of a vector ${\bf x}$ of Lagrange multipliers such that
$$
\kk ^{-1}\cdot \Delta {\bf f}^* = \rig \cdot {\bf x},
$$
and \eqref{eqn:consteq} thus entails
$$
\rigt\cdot\kk\cdot\rig\cdot{\bf x}=\sti\cdot{\bf x}=\Delta {\bf F}^{\text{ext}},
$$
This means that ${\bf x}$ is the displacement vector solution to the elastic problem, and that $\Delta {\bf f}^* = \kk\cdot\rig \cdot {\bf x}$ 
is indeed the corresponding contact force increment vector.

We now derive the explicit formulae for Voigt and Reuss bounds. 

A first step is to exploit the isotropy of the medium, which enables inequalities to be written separately for bulk and shear moduli.
In our stress-controlled approach, $\Delta\ww{\sigma}$ is imposed, minimum values in~\eqref{eqn:min1} and 
\eqref{eqn:min2} are 
$$
\ba
W_1^* &= -\frac{\Omega_0}{2}\Delta \ww{\sigma} : \ww{\ww{C}}^{-1} :\Delta \ww{\sigma}\\
W_2^* &= -W_1^*
\ea
$$
The values obtained with trial solutions for displacements or contact force increments can then be regarded
as estimates of those quadratic expressions in $\Delta\ww{\sigma}$, 
and hence provide estimates of the corresponding compliance matrix $\ww{\ww{C}}^{-1}$.
The meaning of $\ww{\ww{C}}^{-1}$, in a finite sample, is specific to the choice of particular boundary conditions. 
In the large sample limit, it is assumed to
satisfy the symmetry properties of the medium, which is statistically isotropic in the numerical studies reported here. 
Moreover, it is also expected to approach the
macroscopic compliance matrix, whatever the particular choice of boundary conditions. If, as in our numerical study, 
we restrict $\Delta\ww{\sigma}$ to a diagonal form, and
hence regard it as a vector with three coordinates $\sigma_\alpha,\ \alpha=1,2,3$, $\ww{\ww{C}}^{-1}$ is a matrix $\ww{S}$ of the form
\be
\ww{S}=\bma
S_{11}&S_{12}&S_{12}\\ S_{12}&S_{11}&S_{12}\\S_{12}&S_{12}&S_{11}
\ema
,
\label{eqn:defms}
\ee
with, due to isotropy, 
$$
\ba
S_{11}&= \frac{1}{9B}+\frac{1}{3G}\\
S_{12}&= \frac{1}{9B}-\frac{1}{6G}.
\ea
$$

When $\Delta\ww{\sigma}$ is an isotropic pressure increment $\Delta P$, one has 
$$
W_2^*=-W_1^*= \frac{\Omega _0(\Delta P)^2}{3B},
$$
whence an upper bound to $B$ with an estimate of $W_1^*$, and a lower bound with an estimate of $W_2^*$.

When  $\Delta\ww{\sigma}$ is of the form $(q,-q,0)$, then
\be
W_2^*=-W_1^*= \frac{\Omega _0 q^2}{2G},
\label{eqn:w2iso}
\ee
whence an upper bound to $G$ with an estimate of $W_1^*$, and a lower bound with an estimate of $W_2^*$.

The Voigt bounds on $B$ and $G$ are based on trial displacements defined as (see Eqn.~\ref{eqn:defu}) 
\be
%\ba
%\tilde {\bf u}_i &= 0\\
%-\ww{\epsilon}\cdot{\bf R}_i\\
%\Delta \theta_i &= 0
%\vec{\omega},
{\bf U} = \left( ({\bf 0},{\bf 0})_{1\le i\le n},(\epsilon_{\alpha})_{1\le \alpha\le 3} \right)
%\ea
\label{eqn:utrial}
\ee
The choice of $\ww{\epsilon}$ should then minimize~\eqref{eqn:min1}, restricted to displacements of this particular form, \emph{i.e.,}
$$
W_1 (\vec{\epsilon}) = 
\frac{\Omega_0}{2}\vec{\epsilon}\cdot\ww{L}\cdot\vec{\epsilon} - \Omega_0 \vec{\sigma}\cdot\vec{\epsilon},
$$
in which the 3-vector notations for strains and stress increments as defined above are adopted, and $3\times 3$ matrix $\ww{L}$ is a sum over contacts,
to be evaluated below. The best choice for $\vec{\epsilon}$ is
$$
\vec{\epsilon}^*=\ww{L}^{-1}\cdot\vec{\sigma}
$$
The minimum value of $W_1$ then yields
\be
\ww{S}^{\text{Voigt}}=\ww{L}^{-1}.
\label{eqn:estimsv}
\ee
Let us now write down matrix $\ww{L}$ in a sphere packing with our boundary conditions.
We introduce the notations 
$$
\ba
L^N_{ij}&=(R_i+R_j)^2K^N_{ij}\\
L^T_{ij}&=(R_i+R_j)^2K^N_{ij}
\ea
$$ for each contact $i,j$ between spheres of radii $R_i$ and
$R_j$, and neglect $h_{ij}$ in comparison with the radii, as we have been doing throughout this article.
 Then one has, for each pair of indices $\alpha, \beta$, $1\le \alpha, \beta\le 3$ (no sum over repeated indices)
\be
L_{\alpha\beta}=\frac{1}{\Omega _0} \sum_{i<j} \left[ L_{ij}^N (n_{ij}^\alpha)^2 (n_{ij}^\beta)^2 +
L_{ij}^T(\delta_{\alpha\beta}-n_{ij}^\alpha n_{ij}^\beta)n_{ij}^\alpha n_{ij}^\beta \right]
\label{eqn:redmat}
\ee
In \eqref{eqn:redmat}, $\delta_{\alpha\beta}$ is the Kronecker symbol and
index pairs $i<j$ run over the list of force-carrying contacts between grains labelled $i$ and $j$.

In general, it might be favorable to allow for a common spin, \emph{i.e.},  $\Delta \theta _i= \vec{\omega}$, 
for all particles in the choice trial displacements~\eqref{eqn:utrial}.
The optimal choice $\vec{\omega}^*$ only vanishes when the stress tensor and the fabric tensor defined as the average of $\nij\otimes\nij$ over contacts, 
weighted by $L_{ij}^T$, share the same eigenvectors. 
This conclusion,wich  was reached before by Jenkins and La Ragione~\cite{JeLa01}, and independently by Gay and da Silveira~\cite{GdaS04},
on directly estimating the stress increments corresponding to a prescribed
strain, is retrieved here as an illustration of the variational approach.

Returning now to the case of isotropic packings of monodisperse spherical beads of diameter $a$,
the fabric tensor is isotropic, which ensures $\vec{\omega}^*$.  Thus
matrix $\ww{L}$ is the Voigt estimate of $\ww{C}$. To compute its terms in the large system limit for isotropic packings, 
we transform the sums in~\eqref{eqn:redmat} as we did for the evaluation of average stiffness by~\eqref{eqn:avkn}.
Exploiting the independence between stiffness fluctuations and contact orientations, as well as relations
$$
\ave{n_x^4}=\frac{1}{5}\ \ \mbox{and}\ \ \ave{n_x^2n_y^2}=\frac{1}{15}
$$  
for isotropically distributed unit vectors, one gets
$$
\ba
C_{11}^{\text{Voigt}}&=\frac{3^{4/3}}{2}\left(\frac{z\Phi\tilde E}{\pi}\right)^{2/3}\frac{3+2\alpha_T}{15}Z(1/3)P^{1/3}\\
C_{12}^{\text{Voigt}}&=\frac{3^{4/3}}{2}\left(\frac{z\Phi\tilde E}{\pi}\right)^{2/3}\frac{1-\alpha_T}{15}Z(1/3)P^{1/3},
\ea
$$
from which expression~\eqref{eqn:bgvoigt} of $B^{\text{Voigt}}$ and $G^{\text{Voigt}}$ is readily derived, since $B = (C_{11}+2C_{12})/3$ and $G=(C_{11}-C_{12})/2$.

Finally, to establish the Reuss lower bound for $B$, we choose an isotropic stress increment $\Delta \vec{\sigma}= (\Delta P, \Delta P, \Delta P)$ and
evaluate $W_2$ for a trial set of contact force increments chosen, in any sample in equilibrium under pressure $P$, as 
\be
\Delta {\bf f}_{ij} = \frac{\Delta P}{P} {\bf f}_{ij},
\label{eqn:dftrial}
\ee
in contact $i,j$, initially carrying force ${\bf f}_{ij}$.
Such force increments balance the load increase $\Delta P$ by linearity of equilibrium relation~\eqref{eqn:consteq}. The resulting value of $W_2$,
$$
W_2 = \left(\frac{\Delta P}{P}\right)^2 \frac {1}{2\Omega_0} \sum_{i<j} \frac{N_{ij}^2}{K_{ij}^N}+  \frac{{\bf T}_{ij}^2}{K_{ij}^T}
$$ 
is quadratic
in $\Delta P$, and yields $B^{\text{Reuss}}$, as written in~\eqref{eqn:encB}, on identifying the optimal $W_2$ value with the macroscopic energy:
$$
W_2^*= \frac{\Omega_0 (\Delta P)^2}{3B},
$$
once the sum is transformed by the same procedures as in the evaluation of the average stiffness $\ave{K_N}$ in~\eqref{eqn:avkn},
using the definition of $\tilde Z(5/3)$ in~\eqref{eqn:tilz}.

No such trial vector of contact force increments as~\eqref{eqn:dftrial} 
is readily available when the applied stress increment is not proportional to the preexisting stress, which
is isotropic in the present study. In general, in anisotropic stress states, a similar Reuss-type approach is expected to provide a lower bound 
estimate for a certain combination of elastic moduli, which expresses the response to proportional load increases.

Finally, let us recall that similar minimization properties as~\eqref{eqn:min1} and~\eqref{eqn:min2} hold in a strain-controlled approach.
If strains $\ww{\epsilon}$ are imposed, then displacement vector ${\bf U}$, which
is constrained to correspond to  $\ww{\epsilon}$ (this sets the values of its three last coordinates with our choice of boundary conditions) should minimize:
$$
\EE_1 ({\bf U}) = \frac{1}{2} {\bf U}\cdot\sti\cdot{\bf U}
$$
while the contact force increments, vector $\Delta {\bf f}$, should equilibrate each grain and minimize
$$
\EE_2(\Delta {\bf f})=\frac{1}{2} \Delta {\bf f}\cdot \kk ^{-1}\cdot\Delta {\bf f} - \Omega\Delta \ww{\sigma}:\ww{\epsilon},
$$
in which the stress increment $\ww{\Delta \sigma}$ is directly given by $\Delta {\bf f}$, just like
stress components relate to contact forces in~\eqref{eqn:stress}.

It is easy to check that the strain-controlled approach yields exactly the same Voigt and Reuss bounds as the stress-controlled one.

%\bibliography{../granu}
\end{document}